\documentclass{mn2e}

\usepackage{graphicx,amsmath,amssymb,subfigure}

\def\simgt{\mathrel{\lower0.6ex\hbox{$\buildrel {\textstyle >}
 \over {\scriptstyle \sim}$}}}
\def\simlt{\mathrel{\lower0.6ex\hbox{$\buildrel {\textstyle <}
 \over {\scriptstyle \sim}$}}}

\newcommand{\Lsolar}{\mbox{\,$\rm L_{\odot}$}}        

\newcommand{\ang}{\mbox{$\rm \AA$}}

\begin{document}

\title[The Fundamental Plane of radio galaxies]
{The evolution of the Fundamental Plane of radio galaxies from $z \sim 0.5$ to the present day}

\author[P.D.~Herbert et al.]
{Peter D. Herbert$^{1}$\thanks{Email: p.d.herbert@herts.ac.uk}, Matt J. Jarvis$^{1}$, Chris J. Willott$^{2}$, Ross J. McLure$^{3}$, \and
Ewan Mitchell$^{4}$, Steve Rawlings$^{4}$, Gary J. Hill$^{5}$ and James S. Dunlop$^{3}$ \\
\footnotesize
$^{1}$Centre for Astrophysics Research, Science \& Technology Research Institute,
       University of Hertfordshire, Hatfield, AL10 9AB, UK \\
$^{2}$ Herzberg Institute of Astrophysics, National Research Council,
       5071 West Saanich Rd, Victoria, BC V9E 2E7, Canada\\
$^{3}$SUPA\thanks{Scottish Universities Physics Alliance} Institute
      for Astronomy, University of Edinburgh, Royal Observatory,
      Edinburgh, EH9 3HJ, UK\\ 
$^{4}$Department of Physics, University of Oxford Astrophysics, Keble Road,
       Oxford, OX1 3RH, UK \\
$^{5}$McDonald Observatory, University of Texas at Austin,
       1 University Station C1402, Austin, TX 78712-1083, USA \\
}
\maketitle

\begin{abstract}
We present deep spectroscopic data for a 24-object subsample of
our full 41-object $z \sim 0.5$ radio galaxy sample
in order to investigate the evolution of the Fundamental Plane of
radio galaxies.
We find that the low-luminosity, FRI-type, radio galaxies in our
sample are consistent with the local Fundamental Plane of radio
galaxies defined by Bettoni et al. when corrected for simple passive
evolution of their stellar populations.  However, we find that the
higher luminosity, FRII-type radio galaxies are inconsistent with
the local Fundamental Plane if only passive evolution is considered,
and find evidence for a rotation in the Fundamental Plane at
$z \sim 0.5$ when compared with the local relation.
We show that neither passive evolution, nor a mass-dependent evolution
in the mass-to-light ratio, nor an evolution in the size of the host galaxies
can, by themselves, plausibly explain the observed tilt. However, we
suggest that some combination of all three effects, with size
evolution as the dominant factor, may be sufficient
to explain the difference between the planes.

We also find evidence for a correlation between host galaxy velocity
dispersion and radio luminosity at the 97\% significance level within
our sub-sample, although further observations are required in order
to determine whether this is different for the FRI and FRII radio sources.
Assuming that the $M_{\rm BH} - \sigma$ relation still holds at
$z \sim 0.5$, this implies that radio luminosity scales with black hole mass,
in agreement with previous studies. 
\end{abstract}


\section{Introduction}\label{intro}

A number of tight correlations exist between the central black hole
and the host galaxy of an active galactic nucleus (AGN): for
example, the relation between black-hole mass and stellar
velocity dispersion ($M_{BH}$-$\sigma$) (Gebhardt et al.
2000; Ferrarese \& Merritt 2000; Merritt \& Ferrarese 2001;
G\"ultekin et al. 2009) and the relation between black hole
mass and bulge luminosity (e.g. McLure \& Dunlop 2001; McLure
\& Dunlop 2002; G\"ultekin et al. 2009). The existence of these relationships
suggests that AGNs have an important role to play in the
formation and evolution of galaxies via AGN-driven feedback.
Various mechanisms have been suggested for this feedback,
including a scenario proposed by Silk \& Rees (1998) in which quasar
outflows limit black hole masses, dependent on the depth of the
potential wells of dark matter halos.

AGN-driven feedback is now
incorporated into models of galaxy formation and evolution: for
example, the semi-analytic model of Croton et al. (2006),
where  AGN-driven feedback is separated into ``quasar-mode''
(associated with the efficient accretion of cold gas) and
``radio-mode'' (associated with the less efficient accretion of
warm gas) feedback. Sijacki et al. (2007) also separate AGN
feedback into two modes in their full hydrodynamical model: 
the ``quasar regime'' (corresponding to central black holes
with high accretion rates) and mechanical feedback (corresponding
to central black holes with low accretion rates).
Despite the nomenclature, radio galaxies can exhibit either
``quasar-mode'' or ``radio-mode''/mechanical feedback,
dependent on the accretion onto the central black hole.
There is increasing evidence (e.g. Hardcastle, Evans \& Croston
2007, 2009) that the accretion mode of the AGN has a direct
link with the emission line classification
scheme introduced by Hine \& Longair (1979), where radio
galaxies are classified as either high-excitation (HEGs)
or low-excitation (LEGs).
By studying the host galaxies of radio-loud AGN spanning
luminosities from the radio mode (typically lower radio luminosities)
through to the quasar mode (typically higher radio luminosities)
we should be able to place important constraints on
the validity of the feedback mechanisms in the models of
galaxy formation and evolution. Furthermore, radio galaxies reside
in the most massive galaxies at all epochs (e.g. Jarvis et al. 2001;
Willott et al. 2003) and as such allow us to trace the evolution of the
most massive galaxies across virtually all of cosmic time.

One method of doing this is to study the
evolution of the Fundamental Plane of radio galaxies.
The concept of the Fundamental Plane (Djorgovski \&
Davis 1987; Dressler et al. 1987)
describes the observation that a three-dimensional
representation of $\log R_{e}$ versus
$\log \sigma$ versus $\log I_{e}$ (where $R_{e}$ is the
effective radius, $\sigma$ the velocity dispersion and
$I_{e}$ the effective surface brightness of the galaxy) for local
elliptical and lenticular galaxies yields a plane.
The Fundamental Plane has a lower scatter than the Faber-Jackson 
$L$-$\sigma$ relation (Faber \& Jackson 1976), and thus was
initially of interest for its use as an alternative
distance determination method. However, it was also realised
that the fundamental plane
scalings and small scatter were important for
constraining models of elliptical galaxy formation. We may
describe the fundamental plane as follows:

\begin{equation}\label{equation:fp}
R_{e} \propto \sigma^{\alpha} I_{e}^{- \beta}.
\end{equation}

Use of the virial theorem with constant mass-to-light ratio
permits us to reproduce this relation
with coefficients of $\alpha = 2$ and $\beta = 1$ (Faber et al.
1987). However, the observed fundamental plane does not follow
this virial scaling.  For example, J\o rgensen, Franx \&
Kj\ae rgaard (1996) find $\alpha = 1.24 \pm 0.07$ and $\beta = 0.82 \pm 0.02$. This
``tilt'' in the fundamental plane suggests that the mass-to-light
ratio ({\it M}/{\it L}) varies as a function of galaxy mass
(Robertson et al. 2006).
Faber et al. (1987) also note that deviations from the
fundamental plane can be caused by other changes in 
{\it M}/{\it L}, for example due to stellar metallicity
or age, or the distribution of dark and baryonic matter.
Robertson et al. (2006) comment that each of these effects
could, in principle, introduce a systematic tilt into the
fundamental plane if they vary as a function of galaxy mass.
J\o rgensen et al. (1996) also find that the slope of the
Fundamental Plane is independent of cluster
properties such as cluster velocity dispersion and
gas temperature.


\subsection{The Fundamental Plane of Radio Galaxies \& AGN}\label{agn-fp}

Bettoni et al. (2001) collected photometric
and dynamical data for 73 low redshift ($z < 0.2$) radio
galaxies from a variety of samples. They find that these
local active galaxies lie on the same Fundamental Plane as
the inactive population, with the implication that the gas
accretion of the central black hole does not influence the
global properties of early-type galaxies. Wolf \& Sheinis
(2008) investigate the location on the Fundamental Plane of
ten nearby ($z < 0.3$) luminous quasars. They find that the
radio-loud objects in their sample fall on the upper extreme
region of the Fundamental Plane and have host galaxy
properties similar to those of giant early-type galaxies,
whilst their radio-quiet objects are located in the region
of the Fundamental Plane occupied by normal early-type
galaxies and have host galaxy properties more similar to
intermediate-mass galaxies.


\subsection{Redshift Evolution of the Fundamental Plane}\label{fp-evolution}

Work has also been undertaken to examine the redshift evolution of the
Fundamental Plane. The intermediate redshift Fundamental Plane
($z \sim 0.3$-$0.4$) for cluster (Dokkum \& Franx 1996; Kelson et al. 2000)
and field (Treu et al. 2001) early-type galaxies is found to be similar
to that observed locally. Treu et al. (2001) find an offset (from the
local relation) in their intermediate redshift Fundamental Plane of
early-type field galaxies. This offset increases with redshift and is
attributed to an increase in effective brightness of these galaxies with
respect to a local sample. Woo et al. (2004) study the Fundamental Plane
of the host galaxies of 15 AGN out to a redshift of $z \sim 0.34$.
They find that the host galaxies of these intermediate redshift
BL Lac objects and radio galaxies lie on the same Fundamental Plane as
normal local early type galaxies and the local radio galaxy sample of
Bettoni et al. (2001), although their higher redshift objects possess
an offset in the same fashion as the one obeserved by Treu et al. (2001).
Woo et al. (2004) also observe an increase in the mass-to-light ratio
of their galaxies of $\sim$40\% since $z \sim 0.3$, also in agreement
with the findings for normal early-type galaxies.

At higher redshifts ($z \sim 0.8$-$1.3$) various authors
(di Serego Alighieri et al. 2005; Treu et al. 2005a; J\o rgensen et al. 2006;
J\o rgensen et al. 2007; Fritz \& Ziegler 2009; Fritz et al. 2009; Fritz,
J\o rgensen \& Schiavon 2010) find not only an offset but also a 
rotation in the Fundamental Plane when compared to the local relation.
For example, 
J\o rgensen et al. (2007) find a slope described by $\alpha = 0.60 \pm 0.22$
and $\beta = 0.70 \pm 0.06$ for galaxies in their two $z \sim 0.8$ clusters
whilst di Serego Alighieri et al. (2005) find a slope described by
$\alpha = 0.88 \pm 0.16$ and $\beta = 0.63 \pm 0.04$ for their $0.88 \leq z \leq 1.3$
early-type galaxies. The latter authors show at the 90\% confidence
level that the Fundamental Plane rotates with redshift. This is
interpreted as a mass-dependent evolution of the mass-to-light ratio,
where the evolution is faster for less massive galaxies.
Fritz et al. (2010) study field-type early galaxies and demonstrate
that the rotation of the Fundamental Plane appears to be independent
of the environment of the galaxies. There is thus an emerging
picture in which variation in the mass-to-light ratio causes a
redshift-dependent offset in the Fundamental Plane (see also van Dokkum
\& Stanford 2003; van der Wel et al. 2005). This variation
is mass-dependent, giving rise to the redshift-dependent rotation
observed in the Fundamental Plane. In this picture it is galaxy
mass, rather than environment, which plays the major role in
determining galaxy evolution (Treu et al. 2005b), and less massive
distant galaxies have much younger stellar populations than more massive
galaxies [consistent with the downsizing scenario of Cowie et al. (1996)
in which the most massive galaxies form earliest]. It should, however,
be noted that there is not universal agreement on this picture:
Gebhardt et al. (2003) study the Fundamental Plane of 36 field galaxies
(21 early-type and 15 disc galaxies) in the redshift range 0.3-1.0
(with a median redshift $z = 0.8$) and find no difference in the slope
of the plane when compared to the local relation.

Robertson et al. (2006) employ a multiphase interstellar medium model
to investigate the contribution made by various different
factors to the tilt of the fundamental plane. They find that merging
gas-rich disc galaxy models (with gas fractions $f_{gas} > 0.3$)
including the effects of gas dissipation
(cooling, star formation and supernova feedback) yields a scaling of
the Fundamental Plane which is similar to the observed infrared
Fundamental Plane of Pahre, Djorgovski \& de Carvalho (1998).
Changing the redshift of the progenitor galaxies within
the range $z =$ 0-6 has little effect on the plane produced.
The inclusion of gas dissipation
has the effect of decreasing the total-to-stellar mass ratio
($M_{total}$/$M_{\star}$) in the central regions of the galaxies.
This decrease varies as a function of galaxy mass (with a greater
decrease in lower mass systems) thus producing the required tilt
in the Fundamental Plane. Re-merging the remnants of these gas-rich
galaxy mergers leaves the Fundamental Plane largely intact.
Including the effects of feedback from accreting supermassive black
holes in their models makes little difference to the Fundamental
Plane inhabited by the merger remnants. However, use of these
full-model simulations allow Robertson et al. (2005) to reproduce
the power-law scaling of the $M_{BH}$-$\sigma$ relation between
$z = 6$ and $z = 0$. The remnant remergers of Robertson et al. (2006)
preserve the  $M_{BH}$-$\sigma$ relation (albeit with increased
scatter).

\subsection{The Fundamental Plane at $z \sim 0.5$}\label{this-paper}

In this paper we present spectroscopic data for a sample of
radio galaxies at  $z \sim 0.5$.  We use a direct
spectral fitting procedure to obtain the velocity dispersions
of our galaxies. The effective radii and surface
brightnesses for our sample have already been determined
and published (see  McLure et al. 2004).
Combining our velocity dispersion results with this data
thus permits us to study the Fundamental Plane at
$z \sim 0.5$ and draw comparisons with the well
documented local relation.
Our results also allow us to investigate any possible
correlation between the low-frequency radio luminosity at
151~MHz, $L_{151}$, and the velocity dispersion. This enables us to investigate
(via the $M_{BH}$-$\sigma$ relation) the possible link between radio
luminosity and black hole mass in radio-loud AGN (e.g. Nelson
\& Whittle 1996; McLure \& Jarvis 2004; Jarvis \& McLure 2006).

We begin in Section \ref{thesample} of the paper by
describing our sample, before detailing our observations
and data reduction in Section \ref{observations}. Section
\ref{spectralfitting} describes our spectral fitting and
aperture correction techniques, followed by a description
of our Fundamental Plane fitting in Section \ref{planefitting}.
We present and discuss our findings
in Section \ref{discussion} before concluding in Section
\ref{conclusions}. Unless otherwise stated, throughout
the paper we assume a standard cosmology in which
$H_{0}$ = 70 km s$^{-1}$, $\Omega_{M}$ = 0.3 and
$\Omega_{\Lambda}$ = 0.7.

\begin{table*}
\centering
\caption{\label{table:master}The $z \sim 0.5$ sample.  Column 1 lists the
radio galaxy names and columns 2 and 3 list the J2000 source
coordinates.  Column 4 lists the object redshifts, column 5 the
logarithm of the 151-MHz luminosities in units of WHz$^{-1}$sr$^{-1}$ and
column 6 the effective radii in kiloparsecs from McLure et al. (2004).
The effective radii are derived from fits in which the S\'ersic
parameter is allowed to vary.
Column 7 lists the mean \textit{R}-band surface brightness within the
effective radii in mag arcsec$^{−2}$, also from McLure et al. (2004).
Column 8 lists the aperture corrected velocity dispersion of our best
fit and column 9 the reduced $\chi^{2}$ of this fit, whilst column 10
gives the telescope used for the spectroscopic observations.}
\begin{tabular*}{16.0cm}{lccccccccc}
\hline
Source & RA & Dec & z & $L_{151}$ & $r_{e}$ & $\langle \mu \rangle_{e}$ & $\sigma$ & Reduced $\chi^{2}$ & Telescope \\
\hline
3C16 & 00 37 45.39 & +13 20 09.6 & 0.405 & 26.82 & 22.9 & 21.75 & - & - & WHT \\[0.1cm]
3C19 & 00 40 55.01 & +33 10 07.3 & 0.482 & 26.96 & 30.9 & 21.86 & $359^{+52}_{-44}$ & 1.16 & WHT \\[0.1cm]
3C46 & 01 35 28.47 & +37 54 05.7 & 0.437 & 26.84 & 15.8 & 20.78 & $345^{+39}_{-33}$ & 1.34 & WHT \\[0.1cm]
3C172 & 07 02 08.32 & +25 13 53.9 & 0.519 & 27.17 & 12.6 & 20.65 & $328^{+34}_{-29}$ & 1.10 & Gemini \\[0.1cm]
3C200 & 08 27 25.38 & +29 18 45.5 & 0.458 & 26.92 & 13.2 & 20.15 & $343^{+42}_{-37}$ & 1.39 & Gemini \\[0.1cm]
3C244.1 & 10 33 33.97 & +58 14 35.8 & 0.428 & 27.10 & 15.8 & 20.76 & $234^{+21}_{-19}$ & 1.31 & WHT \\[0.1cm]
3C295 & 14 11 20.65 & +52 12 09.0 & 0.464 & 27.68 & 29.5 & 21.03 & $337^{+22}_{-21}$ & 0.99 & Gemini \\[0.1cm]
3C341 & 16 28 04.04 & +27 41 39.3 & 0.448 & 26.88 & 16.6 & 21.08 & $468^{+42}_{-39}$ & 1.78 & WHT \\[0.1cm]
3C427.1 & 21 04 07.07 & +76 33 10.8 & 0.572 & 27.53 & 18.2 & 20.68 & - & - & Gemini \\[0.1cm]
3C457 & 23 12 07.57 & +18 45 41.4 & 0.428 & 27.00 & 14.1 & 20.40 & - & - & WHT \\[0.1cm]
\\
6C0825+3407 & 08 25 14.59 & +34 07 16.8 & 0.406 & 26.09 & 14.5 & 21.12 & - & - & WHT \\[0.1cm]
6C0850+3747 & 08 50 24.77 & +37 47 09.1 & 0.407 & 26.15 & 12.6 & 20.41 & - & - & WHT \\[0.1cm]
6C0857+3945 & 08 57 43.56 & +39 45 29.0 & 0.528 & 26.34 & 11.7 & 20.15 & $272^{+23}_{-20}$ & 1.13 & Gemini \\[0.1cm]
6C1303+3756 & 13 03 44.26 & +37 56 15.2 & 0.470 & 26.29 & 12.3 & 20.65 & $390^{+18}_{-18}$ & 3.24 & Gemini \\[0.1cm]
\\
7C0213+3418 & 02 13 28.39 & +34 18 30.6 & 0.465 & 25.66 & 6.3 & 19.47 & $267^{+24}_{-21}$ & 1.26 & WHT \\[0.1cm]
7C0219+3423 & 02 19 37.83 & +34 23 11.2 & 0.595 & 25.98 & 10.9 & 20.45 & - & - & Gemini \\[0.1cm]
7C0810+2650 & 08 10 26.10 & +26 50 49.1 & 0.435 & 25.58 & 14.8 & 21.16 & $121^{+19}_{-17}$ & 1.27 & WHT \\[0.1cm]
7C1731+6638 & 17 31 43.84 & +66 38 56.7 & 0.562 & 25.52 & 6.0 & 20.47 & $114^{+27}_{-24}$ & 1.12 & Gemini \\[0.1cm]
7C1806+6831 & 18 06 50.16 & +68 31 41.9 & 0.580 & 26.36 & 14.8 & 20.54 & $302^{+18}_{-15}$ & 1.15 & Gemini \\[0.1cm]
\\
TOOT0009+3523 & 00 09 46.90 & +35 23 45.1 & 0.439 & 24.79 & 12.6 & 20.48 & $272^{+23}_{-20}$ & 1.35 & WHT \\[0.1cm]
TOOT0018+3510 & 00 18 53.93 & +35 10 12.1 & 0.416 & 25.16 & 19.5 & 21.21 & $232^{+17}_{-16}$ & 1.29 & WHT \\[0.1cm]
TOOT1626+4523 & 16 26 48.50 & +45 23 42.6 & 0.458 & 25.03 & 7.4 & 19.54 & $241^{+23}_{-20}$ & 1.24 & WHT \\[0.1cm]
TOOT1630+4534 & 16 30 32.80 & +45 34 26.0 & 0.493 & 25.17 & 10.2 & 19.71 & $257^{+9}_{-9}$ & 2.10 & WHT \\[0.1cm]
TOOT1648+5040 & 16 48 26.19 & +50 40 58.0 & 0.420 & 25.12 & 9.5 & 19.71 & $270^{+18}_{-17}$ & 1.18 & WHT \\[0.1cm]
\hline
\end{tabular*}
\end{table*}

\section{The Sample}\label{thesample}

Our full 41-object $z \sim 0.5$ radio-galaxy sample (the ZP5 sample)
consists of all of the narrow-line radio galaxies in the redshift
interval $0.4 < z < 0.6$ from four complete, low-frequency
selected radio surveys; 3CRR (Laing, Riley \& Longair 1983), 6CE
(Eales et al. 1997; Rawlings, Eales \& Lacy 2001), 7CRS (Lacy et al.
1999; Willott et al. 2003) and TexOx-1000 (``TOOT''; Hill \& Rawlings 2003;
Vardoulaki et al. 2010).
The choice of $0.4 < z < 0.6$ was motivated by several factors.
In particular, a sample at $z \sim 0.5$ provides a full three
decades in radio luminosity (including some of the most radio
luminous objects in existence). It therefore allows us to construct
subsamples of objects, divided by radio luminosity, containing sufficient
numbers of objects to investigate correlations of various
properties with radio luminosity.
In addition, the use of $0.4 < z < 0.6$ ensures that satisfactory
data can be obtained (both Hubble Space Telescope imaging and
spectroscopic data) without excessively long integration times.
Full details of the sample can be found in McLure et al. (2004) and
Herbert et al. (2010).

\begin{figure}
\centering
\includegraphics[width=0.45\textwidth]{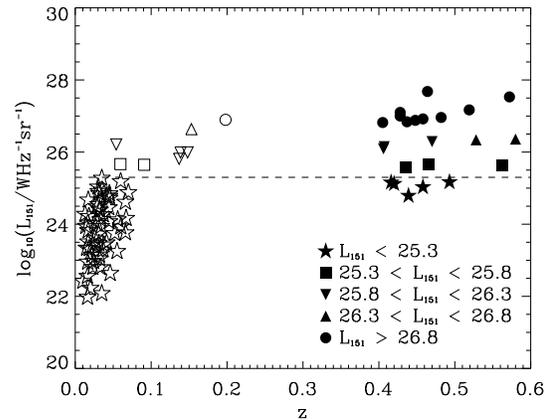}
\caption{\label{L151-z}The logarithm of the 151-MHz radio luminosity,
$L_{151}$, versus redshift for the ZP5 ($z \sim 0.5$) objects in this paper
(filled symbols). We also use open symbols to show the local radio
galaxies from Bettoni et al. (2001; $z \sim 0.03$).
In each case the symbols correspond to the 151-MHz radio luminosities.
Objects with $L_{151} < 10^{25.3}$ W~Hz$^{-1}$~sr$^{-1}$
are shown as stars. Objects with $L_{151}$ between $10^{25.3}$ and
$10^{25.8}$  W~Hz$^{-1}$~sr$^{-1}$ are shown as squares whilst
objects with  $L_{151}$ between $10^{25.8}$ and $10^{26.3}$ 
W~Hz$^{-1}$~sr$^{-1}$ are shown as inverted triangles. Objects with
$L_{151}$ between $10^{26.3}$ and $10^{26.8}$  W~Hz$^{-1}$~sr$^{-1}$
are shown as triangles, and objects with  $L_{151} > 10^{26.8}$
W~Hz$^{-1}$~sr$^{-1}$ are shown as circles.
We do not show five of the objects of Bettoni et al. (0131-360, 0325+024,
PKS 2322-122, 0112+084 and gin116) for which we have been unable to
locate published radio luminosity data.}
\end{figure}

In this paper we use a 24-object subsample of the full ZP5 sample
(see Section \ref{observations} and Table \ref{table:master}).
Spectroscopic observations were made using the William Herschel
Telescope (WHT) or the Gemini North telescope. Objects were
chosen at random from the full sample, aside from the constraint
of $z < 0.5$ for objects observed with the WHT.
In Figure \ref{L151-z} we show the radio luminosity-redshift plane
for the ZP5 objects in this paper.

\section{Observations and Data Reduction}\label{observations}

\subsection{WHT Spectroscopy}\label{spectroscopy:wht}

In August 2002 we used the ISIS spectrograph on the 4.2-metre
William Herschel Telescope
to obtain spectra for 15 of the
objects in the full sample.  The objects were chosen to have
$z < 0.5$ so that satisfactory signal to noise ratios could be
obtained without excessively long integration times. This constraint
aside, the objects were chosen at random from the objects in the
full sample that were visible on the dates of observation.
Details of the objects observed with the WHT can be found in
Table \ref{table:master}. We used the R316R grating on the red arm
together with a 1.0 arcsec slit resulting in a resolution
of 3.7 \AA\ FWHM. This corresponds to a velocity resolution of
FWHM = 185~km~s$^{-1}$ at 6000\AA\  which results in a minimum
measurable velocity dispersion of $\sigma \sim 80$ km s$^{-1}$.
The central wavelength used was dependent on the
redshift such that the spectra contain the Mg$b$ complex to determine
the velocity dispersion. Each object
was observed for a total integration time of 7200s,
split into three 2400s observations to allow the
removal of cosmic rays. The individual spectra were bias subtracted,
illumination and flat-field corrected, and wavelength and flux-calibrated
using standard  {\sc IRAF} routines. Coaddition was done using an average
sigma-clipping routine and the galaxy was extracted using an aperture
defined at the FWHM of the galaxy light profile and at full-width zero
intensity to ensure both high signal-to-noise ratio in the former and
total light spectra for the latter.

\subsection{GMOS Spectroscopy}\label{spectroscopy:gmos}

Between August 2008 and June 2009 we used
the Gemini Multi-Object Spectrograph (GMOS) on the
8.1-metre Gemini North telescope to obtain
spectra for nine of the objects in the full sample. Objects
with $z > 0.5$ were permitted and were chosen at random (given RA
constraints) from the objects in the ZP5 sample left unobserved after the WHT
observations. Details of the objects observed using Gemini
can be found in Table \ref{table:master}.
The B600\_G5303 grating (and subsequently the
new B600\_G5307 grating) was used together with a 0.75 arcsec slit
resulting in a spectral resolution of 4.0 \AA\ FWHM. This corresponds to
a velocity resolution of FWHM = 190~km~s$^{-1}$ at 6200\AA\ which results
in a minimum measureable velocity dispersion of $\sigma \sim 80$ km s$^{-1}$.
Within each observation block, two 600s observations were made
with a central wavelength of 610nm, and a further two 600s 
observations were made using a central wavelength of 605nm.
The use of two
different central wavelengths allows us to cover the gaps
between the CCDs.
Each observation block was repeated four times per source.  Thus
each object was observed for 9600s in total.
The GMOS spectra were reduced (bias subtraction, cosmic ray
subtraction, flat field correction, wavelength calibration,
sky subtraction, flux calibration and spectrum extraction)
using the \textit{gmos} tools in
the \textit{gemini} IRAF package.

\section{Spectral Fitting}\label{spectralfitting}

\subsection{Our Fitting Procedure}\label{fittingprocedure}

In order to determine the velocity dispersion of each
galaxy we perform a  $\chi^{2}$ minimisation, fitting
each galaxy spectrum to a set of templates. The template
set comprises 23 Elodie stellar templates
\footnote{http://atlas.obs-hp.fr/elodie/}
and (following Greene \& Ho 2006) 34 G and K
stars from the old open cluster M67 observed with SDSS.
The templates are broadened using a series of
Gaussians whose width depends on the velocity dispersion
($\sigma$) and wavelength. We exclude from our fit
regions of the galaxy spectra that feature emission
lines ionised by the central engine,
as well
as skylines. The effect of broadening due to the instrumental
resolution is accounted for in our fitting procedure, and the effect is
typically within the uncertainties of our fitted velocity dispersions.

\begin{figure*}
\centering
\caption{\label{Fittings}Our final fits.  The left hand panels show
the galaxy spectrum (top) and the fitted template
(bottom).  A downward shift of 0.3 has
been applied to the template spectra. The shaded areas
indicate regions of the spectra removed from the fitting
due to the presence of emission (rest frame 3859--3880\AA,
3963--3978\AA, 4095--4112\AA, 4332--4378\AA, 4674--4698\AA\ and
4854--4875\AA) or telluric absorption (5570--5585\AA, 6864--6911\AA,
7166--7323\AA\ and 7590--7694\AA) features.
The additional masked regions in 3C172, 6C1303+3756 and 7C1731+6638
were due to artefacts on the chip.
The right hand panels show the $\chi^{2}$ minimisation
as a function of $\sigma$ Aperture corrections have been
applied as described in the text.}
\includegraphics[width=0.95\textwidth]{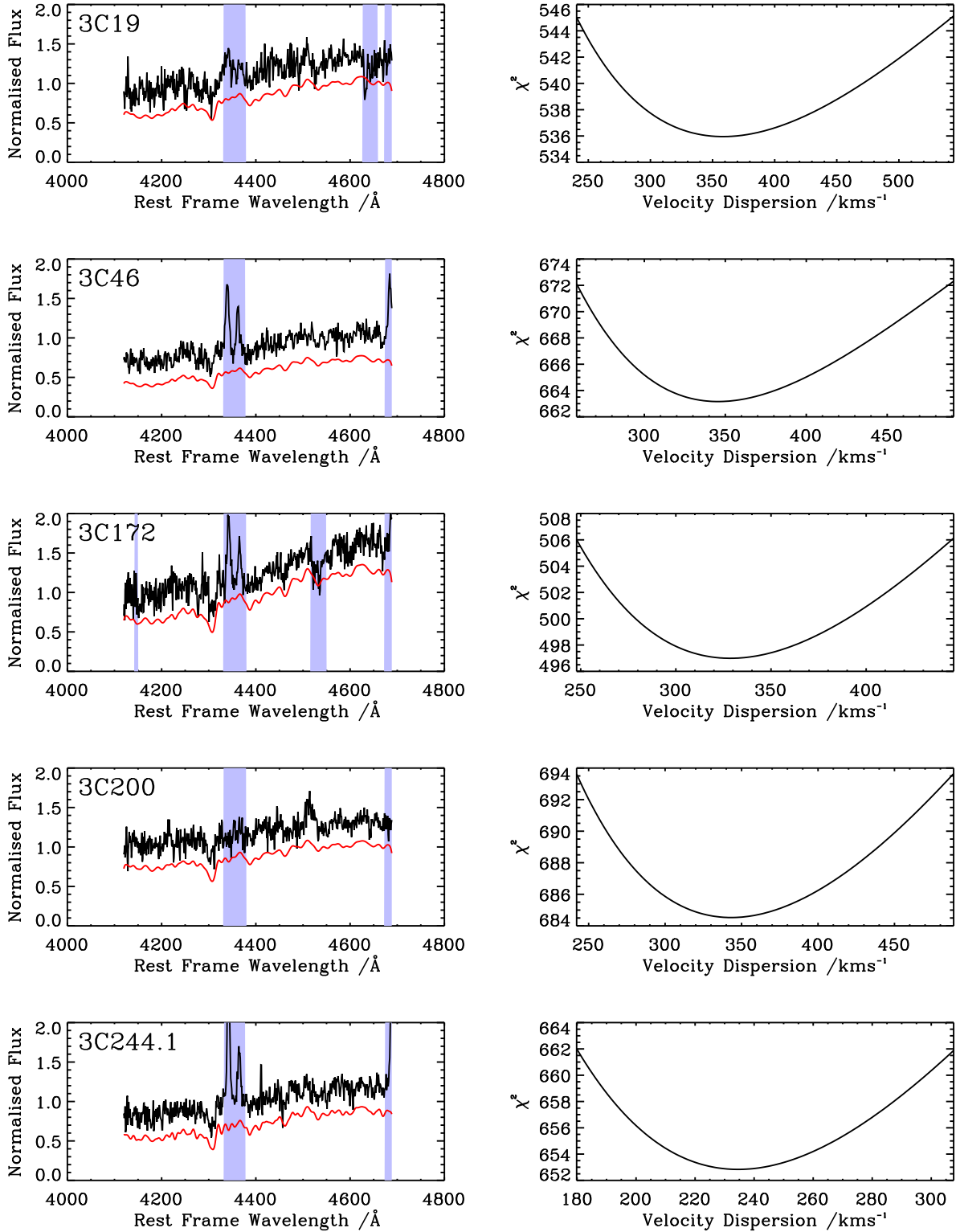}
\end{figure*}

To implement our fitting we fit each galaxy in turn,
using all 57 templates and with the
velocity dispersion incrementing in steps of 1 km s$^{-1}$. At
each step we minimise the reduced $\chi^{2}$ of the
fit using the Amoeba routine in IDL. We then inspect the resulting
matrix of reduced $\chi^{2}$ values in order to find
the minimum reduced $\chi^{2}$ and thus the best
fitting template and velocity dispersion.
Using simulated spectra generated from our stellar template library,
we find that our minimum measureable velocity dispersion is
$\lesssim$ 100~km~s$^{-1}$, well below the expected velocity dispersions
of our radio galaxy hosts.
We are unable to determine reliable measurements of the
velocity dispersion for the cases where the mean signal-to-noise
ratio per resolution element (measured between rest frame 4400\AA\ and 4600\AA\ where
spectral features, in particular the Mg\textit{b} complex,
dominate the fitting) is less than $\sim$10.
On this basis we exclude our fits to 3C16, 3C427.1, 6C0825+3407,
6C0850+3747 and 7C0219+3423. Since this is based merely on the
quality of the data we do not expect this to introduce any notable
selection effects. In addition, we are unable to
obtain a sensible fit in the case of 3C457 due to a combination
of the signal-to-noise ratio and emission lines from the AGN.
Given that this is the only object lacking a fit due to the
AGN, we are confident that this has a negligible effect
on our results. Our final fits to the remaining objects
are shown in Figure \ref{Fittings}.
Spectra for which we were unable to determine a reliable
measurement of the velocity dispersion are shown in the Appendix.

An advantage of explicity looping over the velocity
dispersion is that it allows us to obtain the
standard errors on $\sigma$ from the $\chi^{2}$ distribution.
We convert the $\chi^{2}$ matrix $\left( \chi^{2}_{i} \right)$
(corresponding to the variation of $\chi^{2}$ with
$\sigma$ for the best-fitting template for each
galaxy) into a probability
distribution matrix $\left( P_{i} \right)$ and normalise.
We then inspect the probability distribution function
in order to determine the range in velocity dispersion
which contains 68.27\% of the probability distribution
function and thus obtain the standard errors
on $\sigma$. As a check we note that this procedure
results in the same standard errors as produced
by $\Delta \chi^{2} = 1$.
We note that three objects (3C341, 6C1303+3756 and TOOT1630+4534)
have reduced $\chi^{2}$ values $> 1.5$. The errors on
these objects are therefore undoubtedly underestimated
(see Table \ref{table:master}).

\begin{figure*}
\centering
\includegraphics[width=0.95\textwidth]{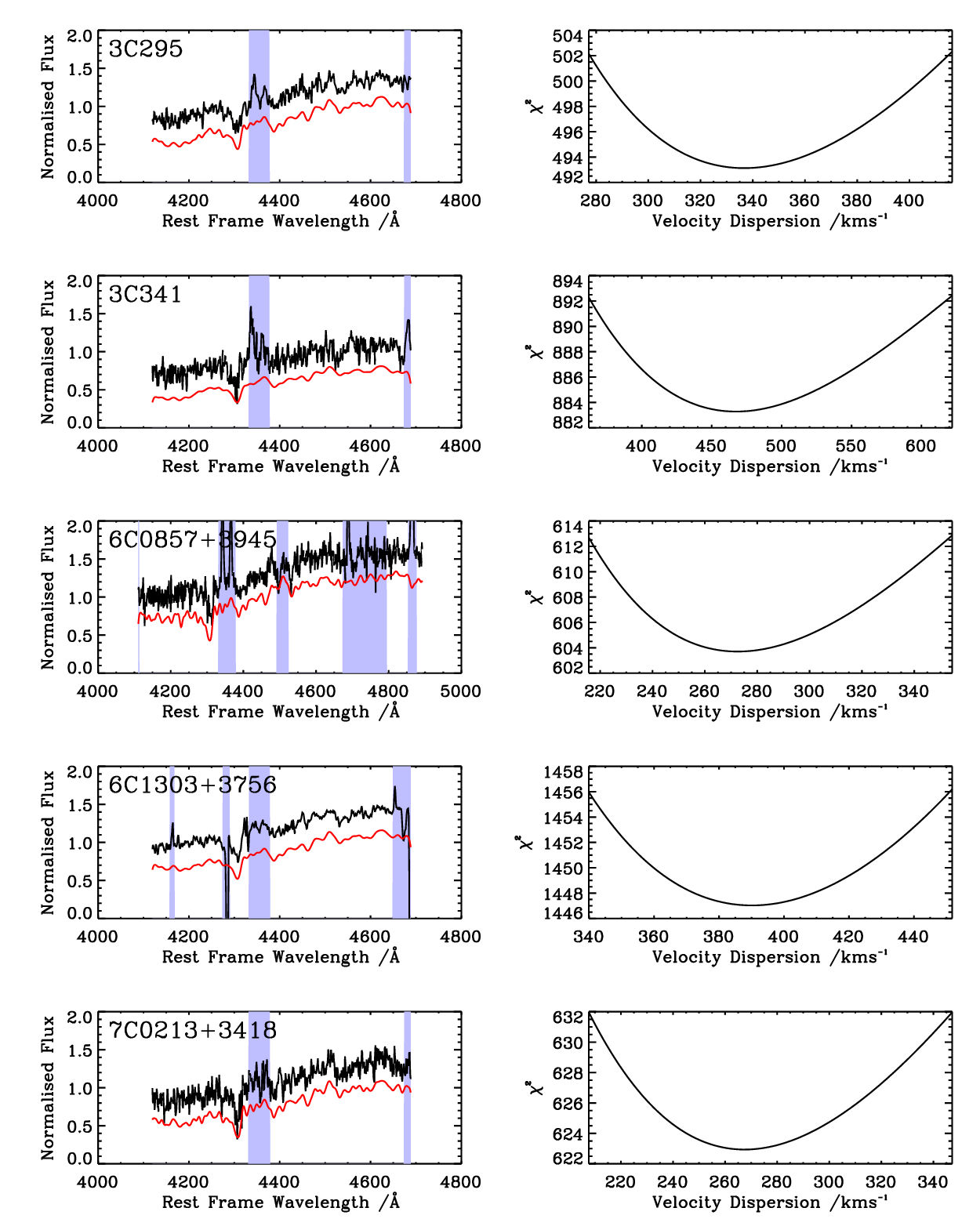}
\vspace{0.4cm}
{\bf Figure \ref{Fittings} continued.}
\end{figure*}

\begin{figure*}
\centering
\includegraphics[width=0.95\textwidth]{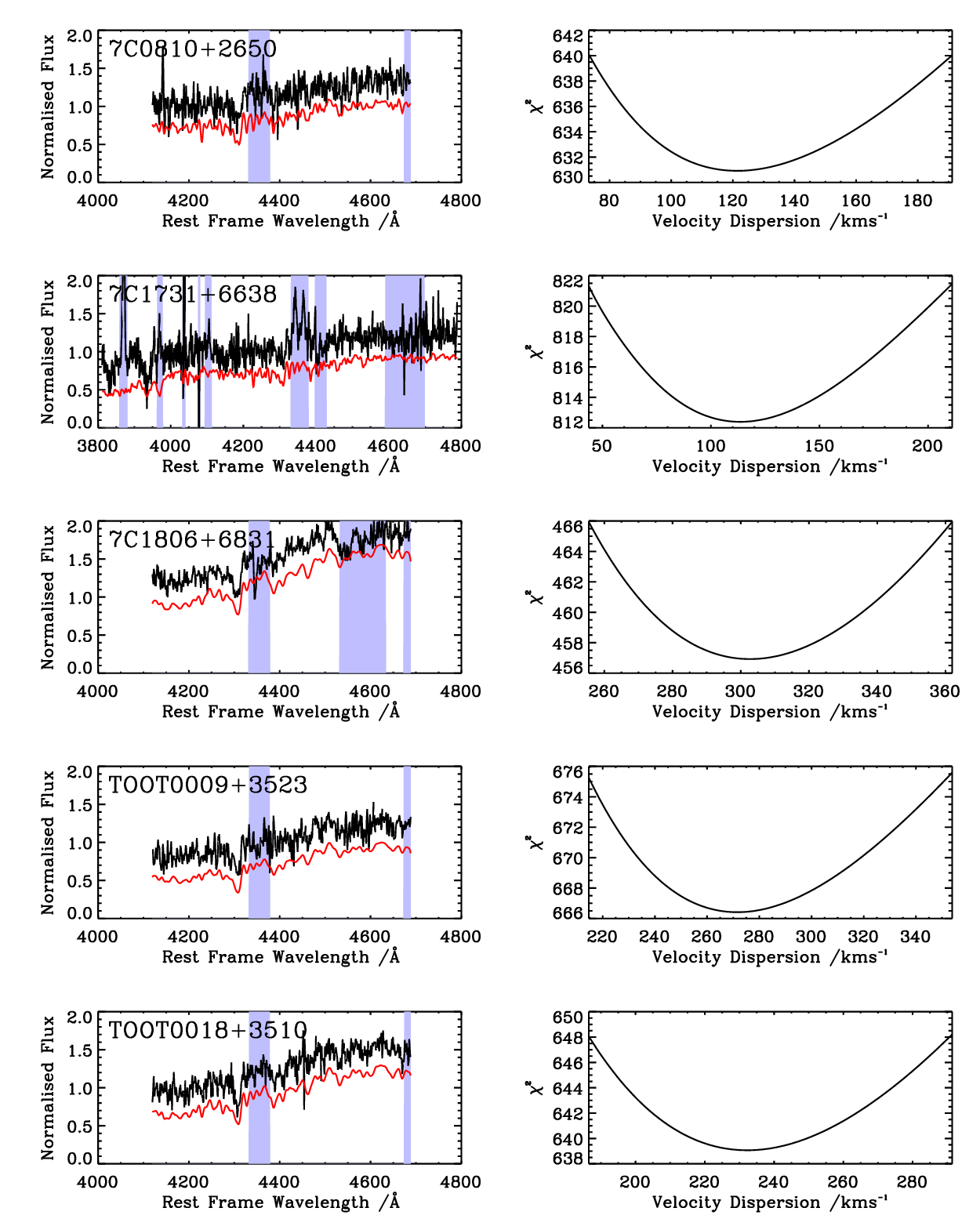}
\vspace{0.4cm}
{\bf Figure \ref{Fittings} continued.}
\end{figure*}

\begin{figure*}
\centering
\includegraphics[width=0.95\textwidth]{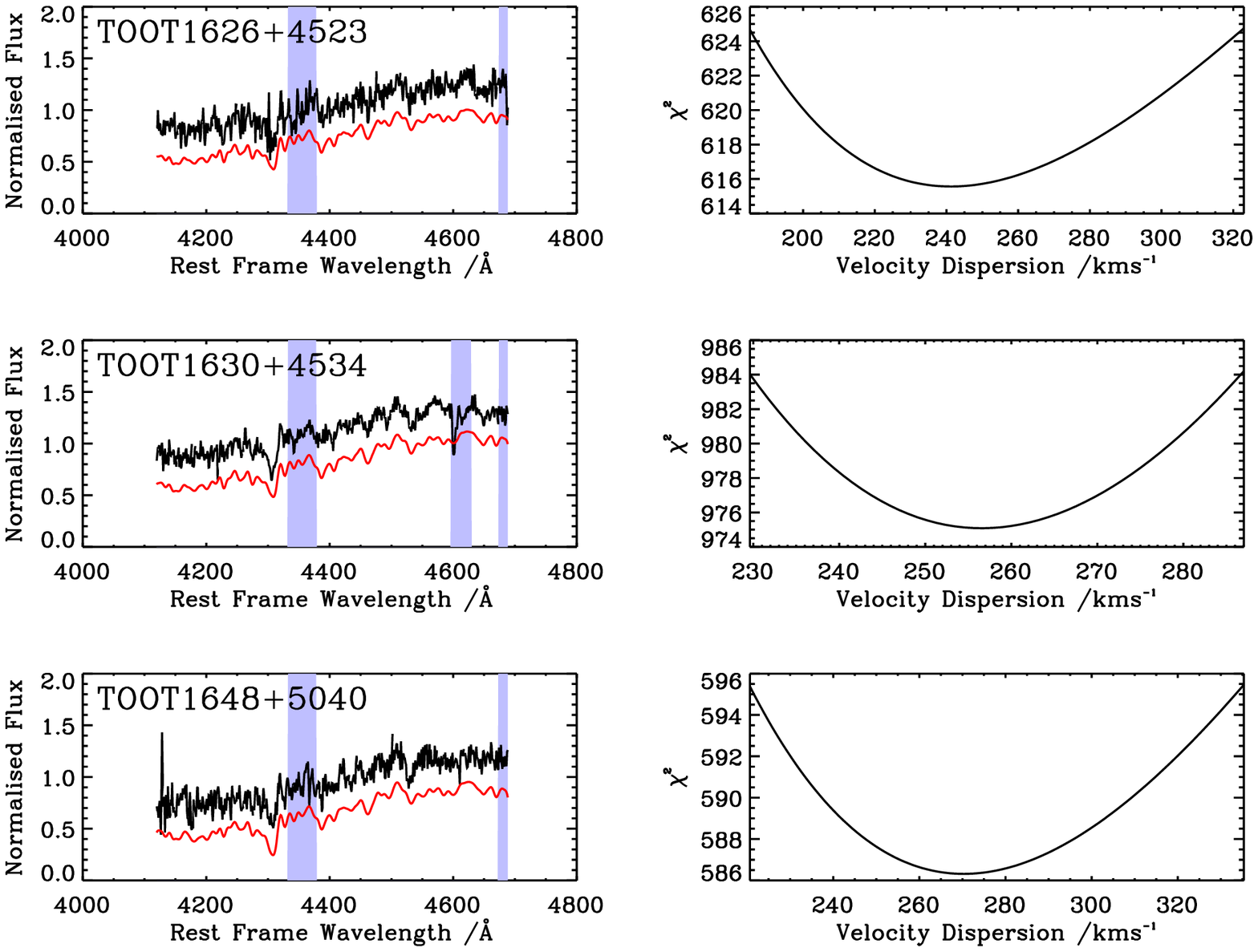}
\vspace{0.4cm}
{\bf Figure \ref{Fittings} continued.}
\end{figure*}


\subsection{Aperture Correction}\label{aperture}

The galaxies in our $z \sim 0.5$ sample are
smaller on the sky than those of a local sample.
Consequently the aperture size for our sample is of order
20 times greater than the aperture size for a local sample.
This has the effect of including more of the galaxy in the
slit, resulting in a lower determined
velocity dispersion than
would be the case for a local sample. Therefore, in order to
be able to compare our results with existing results for
local samples we perform an aperture correction on our
derived velocity dispersion values and their errors.
We perform the aperture correction using the procedure
described by J\o rgensen, Franx \& Kj\ae rgaard (1995),
where our values of $\sigma$ are corrected to a circular
aperture with an aperture diameter of $1.19 h^{-1}$~kpc
(equivalent to 3.4 arcsec projected onto a galaxy in the
Coma cluster). The average correction is 7.1\%.

Our derived velocity dispersion values, including aperture
correction, can be found in Table \ref{table:master}.
Table \ref{table:master} also contains the McLure et al. (2004) values for
the major-axis effective radii, $r_{e}$, and the mean \textit{R}-band surface brightnesses
within $r_{e}$, $\langle \mu \rangle_{e}$, of the galaxies in the sample.  The values
of $\langle \mu \rangle_{e}$ have been corrected for passive evolution as
described in McLure et al. (2004).

\section{Fundamental Plane Fitting}\label{planefitting}

\subsection{Initial Plane Fitting}\label{deltalogfitting}

In order to fit a plane to our data we adopt the
formulation of the fundamental plane used by Robertson
et al. (2006), namely

\begin{equation}\label{equation:planeone}
R_{e} \propto \sigma^{\alpha} I_{e}^{- \beta},
\end{equation}

where $R_{e}$ is the effective radius and $\sigma$
the velocity dispersion as before. $I_{e}$ is the
mean surface brightness within $R_{e}$, in units of
\Lsolar/pc$^{2}$,
and is related to our previous quantity
$\langle \mu \rangle_{e}$:

\begin{equation}\label{equation:muconvert}
\log I_{e} = - 0.4 \left( \langle \mu \rangle_{e} - k \right),
\end{equation}

where for the \textit{R}-band $k = 26.4$
(see J\o rgensen et al. 1996).
In order to fit the fundamental plane
we rewrite Equation \ref{equation:planeone}
in the following form:

\begin{equation}\label{equation:planetwo}
\log R_{e} = \alpha \log \sigma - \beta \log I_{e} - \gamma,
\end{equation}

where $\gamma$ is a constant. For a plane described
by Equation \ref{equation:planetwo} using coordinates
$\left( \mathbf{\log r_{e}}, \mathbf{\log \sigma},
\mathbf{\log I_{e}} \right)$ the residual $\cal{R}$
of point $\left( p_{1}, p_{2}, p_{3} \right)$ 
perpendicular to the plane is given by:

\begin{equation}\label{equation:residuals}
{\cal R} = \frac{1}{\delta} \left( -p_{1} + \alpha
           p_{2} - \beta p_{3} - \gamma \right),
\end{equation}

where

\begin{equation}\label{equation:deltadefn}
\delta = \sqrt{1 + \alpha^{2} + \beta^{2}}.
\end{equation}

We note that 12 of our 18 fitted objects have symmetric errors
on $\log \sigma$. To fit a plane we therefore initially adopt
symmetric errors in $\log r_{e}$, $\log \sigma$, and $\log I_{e}$.
For the six sources which do not have a symmetric error on
$\log \sigma$ we use the mean of
$\left( \log \left( \sigma + \sigma_{upper} \right) - \log \sigma \right)$ and
$\left( \log \sigma - \log \left( \sigma - \sigma_{lower} \right) \right)$, where
$\sigma_{upper}$ and $\sigma_{lower}$ correspond to the upper and lower errors
on $\sigma$ listed in Table \ref{table:master}. The values of
$r_{e}$ and  $I_{e}$ taken from McLure et al. (2004) and used here
have linear errors of 10\%. We use the same procedure as above to estimate
symmetric errors in log space for these quantities. In order to fit a
plane to our results we use a technique similar to the one used by di Serego Alighieri
et al. (2005), where we use a finely sampled grid over $\alpha$, $\beta$
and $\gamma$. At each point we use Equations \ref{equation:residuals} and 
\ref{equation:deltadefn} to calculate the residual perpendicular to the
plane of each of our objects. We use our symmetric errors in log space
in order to determine the error on each residual. We define a
$\chi^{2}$ statistic thus:

\begin{equation}\label{equation:planechisquare}
\chi^{2} = \sum_{i} \frac{{\cal R}_{i}^{2}}{\sigma_{{\cal R}_{i}}^{2}},
\end{equation}

where the summation is over the objects we are fitting the plane to, 
${\cal R}_{i}$ is the residual of object $i$ perpendicular to the plane,
and $\sigma_{{\cal R}_{i}}$ is the standard error on the residual.
Finally we find the values of $\alpha$, $\beta$ and $\gamma$ which
minimise this statistic and thus determine the best fitting plane.
Our best fit fundamental plane can be found in Table \ref{table:rhalfsplanes}.

\begin{table}
\centering
\caption{\label{table:rhalfsplanes}Values of $\alpha$,
$\beta$ and $\gamma$ for a fundamental plane fitting the
ZP5 data and described by the equation
$\log r_{e} = \alpha \log \sigma - \beta \log I_{e} - \gamma$.
The first row gives the results in the case where we assume
symmetric errors in log space (Section \ref{deltalogfitting}).
The second row gives the results from the full Monte Carlo (MC)
fitting using squared residuals (Section \ref{mcsimulations}).
The same procedure, except weighting the minimised residuals
by the error, produces the results listed in the third row.
The results obtained by minimising the sum of the square residuals
and the sum of the absolute residuals without the Monte Carlo
technique are given in the fourth and fifth rows respectively
(Section \ref{mcsimulations}).}
\begin{tabular}{lccc}
\hline
 & $\alpha$ & $\beta$ & $\gamma$ \\
\hline
Symmetric log errors & $0.47_{-0.15}^{+0.16}$ & $0.61_{-0.08}^{+0.12}$ & $-1.3 \pm 0.4$ \\[0.1cm]
MC, squared residuals & $0.52 \pm 0.13$ & $0.63 \pm 0.06$ & $-1.2 \pm 0.4$ \\[0.1cm]
MC, weighted & $0.45 \pm 0.09$ & $0.63 \pm 0.06$ & $-1.4 \pm 0.3$ \\[0.1cm]
Squared residuals & $0.53$ & $0.626$ & $-1.3$ \\[0.1cm]
Absolute residuals & $0.32$ & $0.617$ & $-1.8$ \\
\hline
\end{tabular}
\end{table}

J\o rgensen et al. (1996) and Bettoni et al. (2001) fit a plane by minimising
the sum of the absolute residuals.  However, by using our
$\chi^{2}$-based technique, we can determine the standard errors
on  $\alpha$, $\beta$ and $\gamma$ in a similar fashion to that
employed in Section \ref{fittingprocedure}. We convert our
three-dimensional $\chi^{2}$ matrix into a probability distribution
matrix and normalise. We then determine the volume that 
contains 68.27\% of the probability distribution function and
from this the standard errors on $\alpha$, $\beta$ and $\gamma$.
The reduced-$\chi^{2}$ also provides us with a measure
of the goodness of our fit. Alongside these advantages we do, however, emphasize
that this technique does not make use of the exact errors on $r_{e}$,
$\sigma$ and $I_{e}$ and that small approximations are required.
However, we show in Section \ref{mcsimulations} that the plane we obtain
using this method is consistent with the plane produced using a method
which involves no approximations.

\subsection{Monte Carlo Simulations}\label{mcsimulations}

In order to fit a plane (with errors) to our data, using the actual error
distributions on $\sigma$, we employ a Monte Carlo technique.
For each of our objects we randomly sample 12,000 values from
each of the $r_{e}$, $I_{e}$
and $\sigma$ probability distribution functions (PDFs).
We use sufficient samples to produce smooth final distributions
in $\alpha$, $\beta$ and $\gamma$.
The procedure outlined in section \ref{fittingprocedure}
provides the $\sigma$ PDF.  
For each sample set we again grid over a range of
$\alpha$, $\beta$ and $\gamma$ and obtain the best
fitting values of these parameters for 12,000 simulations
by minimising the sum of the squared residuals according to
equation \ref{equation:residuals}. We obtain
the standard errors by finding the range containing
68.27\% of the PDF for each parameter. Table \ref{table:rhalfsplanes} lists
our fitting results.

Next we repeat this procedure but
rather than minimise just the sum of the
square residuals, we minimise the sum of the square residuals
weighted by the square of the error on the residual
(where we estimate the residual error using the procedure
described in Section \ref{deltalogfitting}). Finally, we
fit a plane to the data by minimising the
sum of the squared residuals, and another by minimising
the sum of the absolute residuals (as per J\o rgensen
et al. 1996; Bettoni et al. 2001), but without any Monte
Carlo sampling in either case. The results from all of
these plane fittings can be found in Table
\ref{table:rhalfsplanes}, and in all cases our Fundamental
Plane parameters are consistent.

\subsection{Local Plane Fitting}\label{localfitting}

In order to compare our $z \sim 0.5$ plane with results for local
active galaxies we use the data presented in Bettoni et al. (2001).
This comprises both new data and data from the literature on their
sample of low redshift radio galaxies. The median redshift of their
sample is 0.03. We note that we use the
same procedure as Bettoni et al. to apply the aperture corrections
and thus our results can be compared directly. The first row of Table
\ref{table:bettoniplanes} contains the published plane fitting
of Bettoni et al. (2001), converted from from their plane
parameterisation to ours using Equation \ref{equation:muconvert}.
The second row of Table \ref{table:bettoniplanes} gives our fit
to the data of Bettoni et al. (2001) using their fitting procedure
(minimising the sum of the absolute residuals perpendicular to the
plane). We exclude from our data set the objects 0053-016 and
0431-134 from Ledlow \& Owen (1995), for which we have been
unable to locate published redshifts.

\begin{table*}
\centering
\caption{\label{table:bettoniplanes} Values of $\alpha$,
$\beta$ and $\gamma$ for a fundamental plane fitting the
data of Bettoni et al. (2001) and described by the equation
$\log r_{e} = \alpha \log \sigma - \beta \log I_{e} - \gamma$.
The first row lists the plane fitting results published by Bettoni
et al. (2001), where we convert from their plane parameterisation to
ours using Equation \ref{equation:muconvert}. The second row lists
our fit to their data obtained by minimising the sum of the
absolute residuals (excluding the two objects discussed in the
text), and the fitting described in the third row is the same
but using the data of Bettoni et al. (2001) corrected into our
cosmology. Row four gives the results of our fit to the
corrected data by minimising the sum of the squared residuals,
and the fifth row gives the results where we assume symmetric
errors in log space and minimise the $\chi^{2}$ statistic.}
\begin{tabular}{lccc}
\hline
 & $\alpha$ & $\beta$ & $\gamma$ \\
\hline
Published & $1.92 \pm 0.15$ & $0.64 \pm 0.03$ & $1.93 \pm 0.37$ \\[0.1cm]
Uncorrected absolute residuals & $1.73$ & $0.709$ & $1.3$ \\[0.1cm]
Absolute residuals & $1.76$ & $0.713$ & $1.5$ \\[0.1cm]
Squared residuals & $1.33$ & $0.655$ & $0.6$ \\[0.1cm]
Symmetric log errors & $2.11_{-0.37}^{+0.42}$ & $0.60 \pm 0.05$ & $2.6_{-0.9}^{+1.0}$ \\[0.1cm]
\hline
\end{tabular}
\end{table*}

In order to compare our plane fitting with the data of Bettoni
et al. (2001) we correct the scalelengths in the data from their
cosmology ($H_{0}$ = 50 km s$^{-1}$, $\Omega_{M}$ = 1 and
$\Omega_{\Lambda}$ = 0) to ours. Once again we exclude the
objects 0053-016 and 0431-134 from Ledlow \& Owen (1995)
due to the lack of redshift information. We fit the Fundamental
Plane to the corrected data by minimising the sum of the
absolute residuals and then repeat, this time minimising the
sum of the squared residuals (see Table \ref{table:bettoniplanes}).
Finally, we use the procedure described in Section
\ref{deltalogfitting} to fit the Fundamental Plane to the corrected data
of Bettoni et al. (2001) using a $\chi^{2}$ fitting statistic,
yielding uncertainties on the plane fitting as well as an indication
of the goodness of fit. Uncertainties on $\sigma$ have been
published for the data presented by Bettoni et al. (2001) and
Smith, Heckman \& Illingworth (1990). For the remainder of the data
we use a typical linear error of 10\%. We also assume linear errors of
10\% on $r_{e}$ and $I_{e}$. The fitting results can be found in
Table \ref{table:bettoniplanes}.

\section{Discussion}\label{discussion}

\subsection{The Local Fundamental Plane}\label{discussion:local-fp}

\begin{figure*}
\centering
\includegraphics[width=0.95\textwidth]{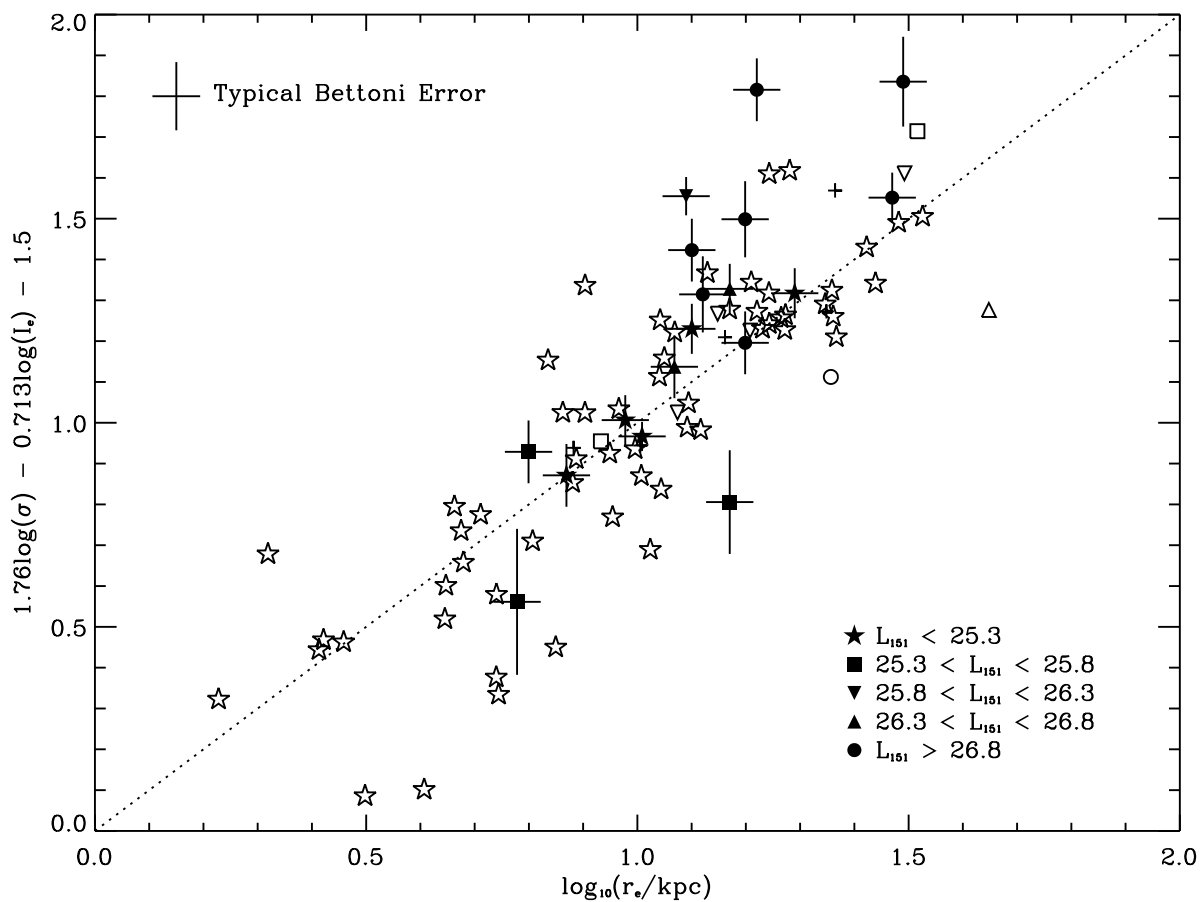}
\caption{\label{absoluteBettoniFP}A 2D projection of the Fundamental
Plane as fitted to the cosmologically corrected data of Bettoni et al.
($z \sim 0.03$) by minimising the sum of the absolute residuals perpendicular to the
plane. The data of Bettoni et al. is shown by open symbols. We use
filled symbols to show the ZP5 data ($z \sim 0.5$), corrected for
passive evolution and aperture corrected as described in the
text.
In each case the symbols correspond to the 151-MHz radio luminosities.
Objects with $L_{151} < 10^{25.3}$ W~Hz$^{-1}$~sr$^{-1}$
are shown as stars. Objects with $L_{151}$ between $10^{25.3}$ and
$10^{25.8}$  W~Hz$^{-1}$~sr$^{-1}$ are shown as squares whilst
objects with  $L_{151}$ between $10^{25.8}$ and $10^{26.3}$ 
W~Hz$^{-1}$~sr$^{-1}$ are shown as inverted triangles. Objects with
$L_{151}$ between $10^{26.3}$ and $10^{26.8}$  W~Hz$^{-1}$~sr$^{-1}$
are shown as triangles, and objects with  $L_{151} > 10^{26.8}$
W~Hz$^{-1}$~sr$^{-1}$ are shown as circles. Objects from Bettoni
et al. for which we have been unable to locate a published radio
luminosity are shown as crosses.
We show the typical assumed error on the data of Bettoni et al.
in the top left-hand corner (see Section \ref{localfitting}).}
\end{figure*}

In Figure \ref{absoluteBettoniFP} we show a two-dimensional
projection of the Fundamental Plane that we fit to the
cosmologically corrected data of Bettoni et al. (2001) using
their fitting procedure (row three of Table
\ref{table:bettoniplanes}). As well as showing the data of
Bettoni et al. in this projection we also show our ZP5 objects with
symbols corresponding to their radio luminosities. The y-axis
uncertainties for the ZP5 objects were determined by assuming errors
as in Section \ref{deltalogfitting} and combining in
quadrature according to the equation of the plane.
It can be seen in Figure \ref{absoluteBettoniFP} that our objects with
radio luminosity $L_{151} < 10^{25.3}$ W~Hz$^{-1}$~sr$^{-1}$ appear to
lie on the local Fundamental Plane. The traditional division between
lower radio luminosity Fanaroff-Riley type I (FRI) and higher radio
luminosity Fanaroff-Riley type II (FRII)
radio galaxies falls at $L_{151} = 10^{25.3}$ W~Hz$^{-1}$~sr$^{-1}$
(Fanaroff \& Riley 1974), and this is also where there appears to be a
difference in the evolution (Clewley \& Jarvis 2004; Sadler et al. 2007),
and thus these objects are probably FRI type radio galaxies, although some
overlap in the populations is inevitable (see Herbert et al. 2010). We note
that the majority of the sample of
Bettoni et al. ($\sim$90\% of the objects which have published radio
luminosities) also have $L_{151} < 10^{25.3}$ W~Hz$^{-1}$~sr$^{-1}$
and are thus classified as FRI type radio galaxies.
We therefore find evidence that the FRI type objects
in our sample can inhabit the local Fundamental Plane of similar
objects simply by passively evolving from $z \sim 0.5$ to the
present day.

Many of the remaining (FRII type, $L_{151} > 10^{25.3}$
W~Hz$^{-1}$~sr$^{-1}$) objects also lie close to the plane and within
the scatter displayed by the data of Bettoni et al. (2001).
However, we note that
our FRII type radio galaxies exhibit a tendency to lie above the
local Fundamental Plane, some by a reasonable distance (although there
are also objects lying a reasonable distance below the plane).
Passive evolution is not sufficient to explain how these objects
would move from their position at $z \sim 0.5$ onto the $z = 0$
Fundamental Plane. The FRII type objects of Bettoni et al. are consistent
with the plane occupied by the rest of the $z \sim 0$ sample,
although there are insufficient objects with high radio
luminosity to study this properly.
Therefore, in contrast to our FRI type objects, we find evidence
that the host galaxies of our $z \sim 0.5$ FRII type objects must undergo evolution
above and beyond their passive evolution to move them onto the
local Fundamental Plane (Section \ref{discussion:evolution}).
We find no clear correlation within the FRII type group
between radio luminosity and the offset from the local Fundamental
Plane, suggesting that this is indeed an evolutionary effect
(rather than one related to the radio luminosity).

We also study the offsets of our ZP5 objects from the local
Fundamental Plane of Figure \ref{absoluteBettoniFP} as a function
of radio morphology and spectral classification. Details of
our classification schemes can be found in Herbert et al. (2010).
We find that those objects with large offsets are predominantly
`Classical Double' (FRII) HEG sources, consistent with the above
result based on radio luminosity and with a similar strength of
trend. It is also worth emphasizing that
we find no link between disturbed morphology from a recent
merger (see McLure et al. 2004) and the size of the offset.

\subsection{The ZP5 Fundamental Plane}\label{discussion:zp5-fp}

\begin{figure*}
\centering
\includegraphics[width=0.95\textwidth]{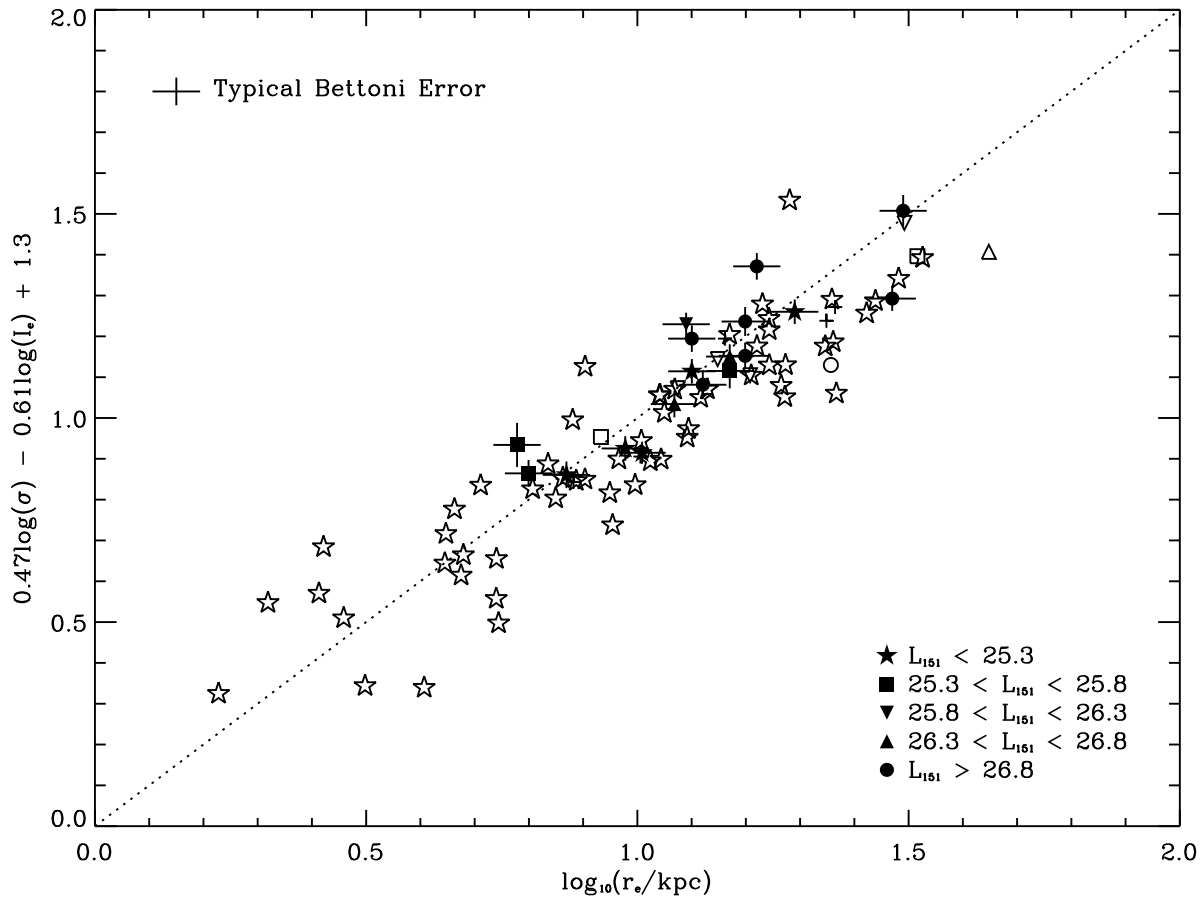}
\caption{\label{deltalogRhalfsFP}A 2D projection of the Fundamental
Plane as fitted to the ZP5 data ($z \sim 0.5$), assuming symmetric errors in log space.
We show the ZP5 objects and the local radio galaxies from
Bettoni et al. (2001; $z \sim 0.03$) in this plane projection, with symbols
as in Figure \ref{absoluteBettoniFP}.}
\end{figure*}

In Figure \ref{deltalogRhalfsFP} we show a two-dimensional
projection of the Fundamental Plane fitting the ZP5 data
assuming symmetric errors in log space (row one
of Table \ref{table:rhalfsplanes}). It is apparent that our
ZP5 objects do inhabit a $z \sim 0.5$ Fundamental Plane,
which is common to both the FRI type and the FRII type
objects in our sample.

The reduced-$\chi^{2}$ of the
ZP5 plane fitting shown in Figure \ref{deltalogRhalfsFP} is 2.9
(tabulated, along with other reduced-$\chi^{2}$ values, in
Table \ref{table:normalredcs}). The reduced-$\chi^{2}$ obtained
when we compare the ZP5 data to the local Fundamental Plane
of Figure \ref{absoluteBettoniFP}
($\alpha = 1.76$, $\beta = 0.713$, $\gamma = 1.5$) is 10.2
(whereas comparing this local plane to the data of Bettoni
et al. 2001 yields a reduced-$\chi^{2}$ of 3.4).
For comparison we also fit a plane to our ZP5 data that follows
the Kormendy ($\mu_{e}$ - $r_{e}$) relation (Kormendy 1977) by
setting $\alpha = 0$ and fitting for $\beta$ and $\gamma$.
Our best fit in this case ($\beta = 0.73$, $\gamma = -2.7$) has
a reduced-$\chi^{2}$ of 5.1.
We find
that our ZP5 fit is a noticeably better fit to the ZP5 data than
the local Fundamental Plane is. It is interesting to note from
row one of Table \ref{table:rhalfsplanes} and row three of Table
\ref{table:bettoniplanes} that this ZP5 fit is significantly
different from the fit to the local data.  The most
prominent difference between the local Fundamental Plane
and the $z \sim 0.5$ Fundamental Plane is that the
$\sigma$ exponent is much smaller for the $z \sim 0.5$ plane.
That is to say, the power law index that relates the
effective radius and the surface brightness to the velocity
dispersion is far smaller at $z \sim 0.5$ than in the local
universe.
The velocity dispersion is related to
the mass ($M$) of the galaxy, and thus we infer that at $z \sim 0.5$
the mass and size of these powerful radio galaxies
are less closely related than in the local universe.
However, despite the smaller $\sigma$ exponent in our ZP5 relation,
a comparison of the
the reduced-$\chi^{2}$ values for our normal ZP5 fit and our
Kormendy fit suggests that our $z \sim 0.5$ Fundamental Plane is
indeed a plane involving $\sigma$ rather than merely a
$\mu_{e}$ - $r_{e}$ relation; a fit with $\alpha = 0$ is noticeably
worse than one where a dependence on $\sigma$ is allowed.

Inspecting Figure \ref{deltalogRhalfsFP} we find that, whilst the
objects of Bettoni et al. are offset with regard to the
ZP5 Fundamental Plane, there nevertheless appears to be a good
agreement between the two data sets. We therefore perform a
plane fitting (assuming symmetric errors in log space) to the
two data sets combined. However, the resulting plane
($\alpha = 1.23$, $\beta = 0.61$, $\gamma = 0.5$) yields
reduced-$\chi^{2}$ values of 4.8, 6.1 and 3.5 when compared
to the combined data, the ZP5 data and the data of Bettoni
et al. respectively. We therefore find that a plane fitted
to the combined data is not acceptable.

In Section \ref{fittingprocedure} we noted that the errors on
$\sigma$ for some of our objects are undoubtedly underestimated.
This contributes in part to the large reduced-$\chi^{2}$ values of
the fits. In order to test whether these underestimated errors are
leading us to reject acceptable plane fittings we repeat our fittings
and comparisons, but requiring a minimum error of 15\% on $r_{e}$,
$\sigma$, and $I_{e}$. The reduced-$\chi^{2}$ values we obtain from
comparing the data to the various plane fittings are shown in
Table \ref{table:minerrorredcs}. We find, with a reduced-$\chi^{2}$
of 1.3, that the local plane (Figure \ref{absoluteBettoniFP})
is an acceptable fit to the data of
Bettoni et al.. We also find that the ZP5 plane (Figure
\ref{deltalogRhalfsFP}) is an acceptable fit to the ZP5 data
(reduced-$\chi^{2}$ of 1.2).
However, we find that the local plane is not an acceptable
fit to the ZP5 data (with a reduced-$\chi^{2}$ of 3.5), and
nor is the ZP5 plane an acceptable fit to the local data
(with a reduced-$\chi^{2}$ of 2.4).
Finally, a plane fitted to the two data sets combined is also
still rejected (reduced-$\chi^{2}$ values of 1.9, 2.6 and 1.3
when compared to the combined data, the ZP5 data and the data
of Bettoni et al. respectively).

Taken altogether, our results therefore suggest that
radio galaxies with lower and higher
radio luminosities may share the same Fundamental Plane, but that there is
substantial evolution in this plane between $z=0$ and $z \sim 0.5$.

\begin{table}
\centering
\caption{\label{table:normalredcs} Reduced-$\chi^{2}$ values.
The first row contains the reduced-$\chi^{2}$ values yielded by
a comparison of the local Fundamental Plane ($\alpha = 1.76$,
$\beta = 0.713$, $\gamma = 1.5$) to our data and that of
Bettoni et al. (2001), whilst
the second row shows the reduced-$\chi^{2}$ values from
comparing the ZP5 plane (row one of Table \ref{table:rhalfsplanes})
to the two data sets.
The third row contains the reduced-$\chi^{2}$ values obtained by
comparing the size-evolved ZP5 plane to the two data sets
(Section \ref{discussion:evolution}), where the ZP5 data is also
ajusted to account for size-evolution.}
\begin{tabular}{lcc}
\hline
& Data of Bettoni et al. & ZP5 Data \\
\hline
Local Plane & 3.4 & 10.2 \\
ZP5 Plane & 5.4 & 2.9 \\
Size-evolved ZP5 Plane & 4.8 & 2.9 \\
\hline
\end{tabular}
\end{table}

\begin{table}
\centering
\caption{\label{table:minerrorredcs} Reduced-$\chi^{2}$ values
where the data used in the plane fittings and for the
comparisons have errors $\geq 15 \%$. The rows and columns are
as in Table \ref{table:normalredcs}.}
\begin{tabular}{lcc}
\hline
& Data of Bettoni et al. & ZP5 Data \\
\hline
Local Plane & 1.3 & 3.5 \\
ZP5 Plane & 2.4 & 1.2 \\
Size-evolved ZP5 Plane & 1.9 & 1.2 \\
\hline
\end{tabular}
\end{table}

\subsection{Evolution of the Host Galaxies?}\label{discussion:evolution}

One explanation for the difference between the local Fundamental Plane of
Bettoni et al. (2001) and our $z \sim 0.5$ Fundamental Plane is
evolution of the host galaxies. We
have already shown in Sections \ref{discussion:local-fp} and
\ref{discussion:zp5-fp} that passive evolution is not sufficient to
explain the difference between the planes, or at least for the
high radio-luminosity population. Other possibilities
include an evolution in the mass-to-light ratio or a size evolution
of the host galaxies. We discuss each of these in turn.

Various authors have found evidence for a mass-dependent evolution
of the mass-to-light ratio (Section \ref{fp-evolution}), and this in
principle could explain the observed rotation between our ZP5 Fundamental
Plane and the local one. 
We use our two plane fittings (row one of Table \ref{table:rhalfsplanes}
and row three of Table \ref{table:bettoniplanes}) to calculate
the mass-dependent evolution of the mass-to-light ratio that would
be required to transform our $z \sim 0.5$ plane into the local one
(assuming $r_{e}$ and $\sigma$ remain constant and expressing $\sigma$
in terms of $r_{e}$ and $M$). However, we find an implausibly large
dependence on the $z \sim 0.5$ luminosity would be required (greater
than 10 orders of magnitude, with a correspondingly small dependence
on the mass), and  we therefore
discount the possibility that this, by itself, could explain the
difference between the two planes.

\begin{figure}
\centering
\includegraphics[width=0.45\textwidth]{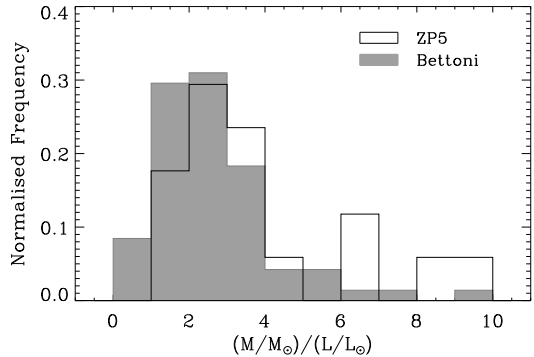}
\caption{\label{mtol-histograms}Histograms of the mass-to-light
ratio distributions for the ZP5 ($z \sim 0.5$) and Bettoni ($z \sim 0.03$) samples.}
\end{figure}

In Figure \ref{mtol-histograms} we show histograms of the mass-to-light
ratio distributions for the ZP5 and Bettoni samples. A two-dimensional
K-S test on the two distributions rejects at the 96\% level the null
hypothesis that the samples are drawn from the same distribution.
We therefore find tentative evidence of an evolution in the mass-to-light ratio
which, whilst not explaining the observed rotation between the $z = 0$ and the
$z \sim 0.5$ planes by itself, may make some contribution to the tilt.
Considering only the FRII type objects in our sample we reject
the null hypothesis that this sample and that of Bettoni et al. are drawn from
the same distribution at the 98\% level. However, we find no clear evidence that
the distribution of mass-to-light ratios is different for our FRI type
objects and the sample of Bettoni et al. (rejecting the null hypothesis at
only the 86\% level), consistent with these objects lying on the local
Fundamental Plane of Figure \ref{absoluteBettoniFP}.

\begin{figure*}
\centering
\includegraphics[width=0.95\textwidth]{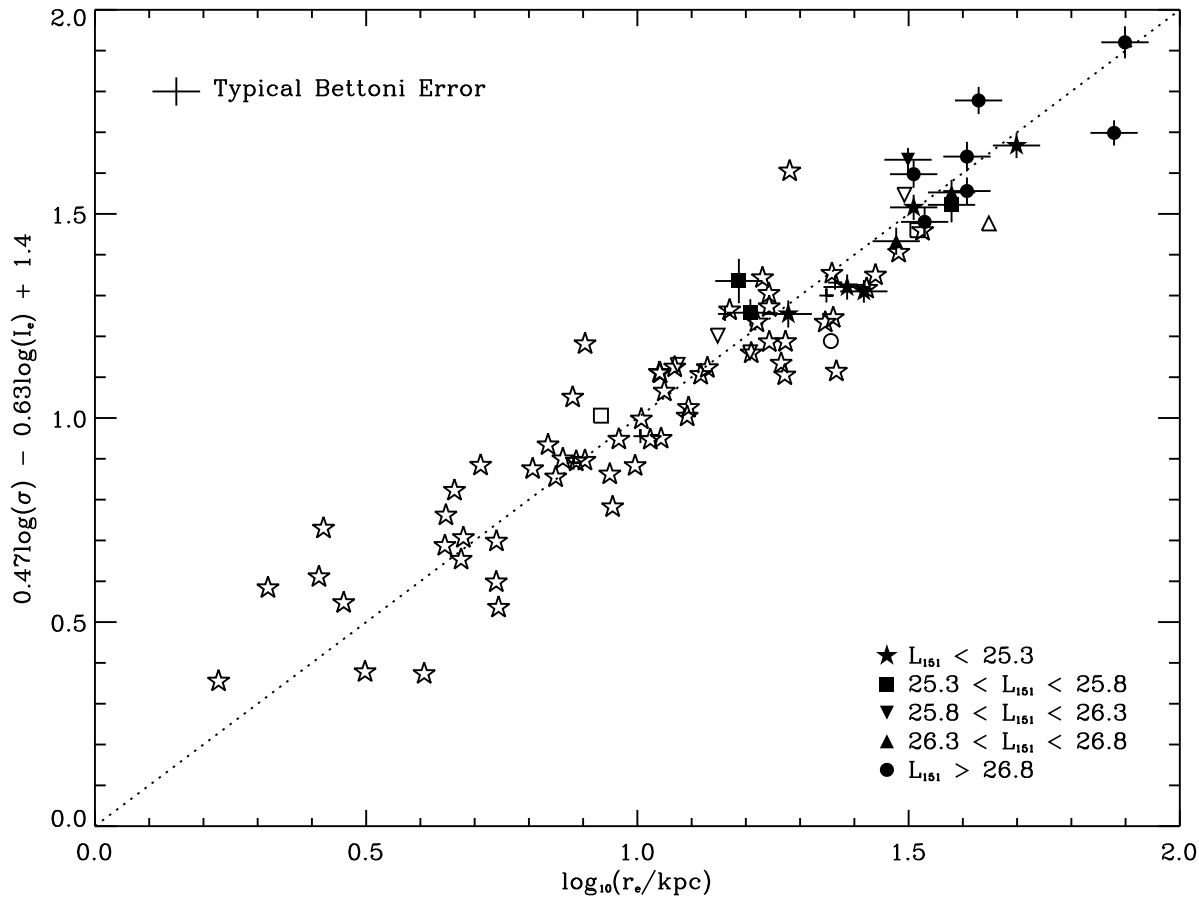}
\caption{\label{evolvedRhalfsFP}A 2D projection of the Fundamental
Plane as fitted to the ZP5 data ($z \sim 0.5$), size-evolved by a factor
$\delta r_{e} = 0.39$.
We show the size-evolved ZP5 objects and the local radio galaxies from
Bettoni et al. (2001; $z \sim 0.03$) in this plane projection, with symbols
as in Figure \ref{absoluteBettoniFP}.}
\end{figure*}

We next turn to consider the possibility of size evolution of the
host galaxies. There exists evidence for the growth of early-type
galaxies from higher redshifts to the present day: for example,
the observational work of van der Wel et al. (2008) and the model of
Hopkins et al. (2010), but see Mancini et al. (2010). We use the size evolution
model of van der Wel et al. (2009), where mergers cause the mass
and effective radius to increase by the same factor. Thus $\sigma$
remains constant (since $\sigma \propto \sqrt{M/R}$), along with the
mass-to-light ratio, whilst  $r_{e}$ and $I_{e}$ vary. We characterise the size evolution using
a factor $\delta r_{e}$ where $r_{e}\left(z = 0.5 \right) = \delta r_{e}
\cdot r_{e}\left(z = 0 \right)$. We investigate what value of $\delta r_{e}$
gives the smallest offset between the ZP5 plane and the local data
by varying $\delta r_{e}$ (and adjusting $r_{e}$ and $I_{e}$ accordingly),
refitting a plane using the procedure described in Section \ref{deltalogfitting},
and minimising the reduced-$\chi^{2}$ of the fit of the local data
of Bettoni et al. (2001) to this plane. For each value of $\delta r_{e}$ we
adjust the amount of passive evolution such that our ZP5 objects
still satisfy the Kormendy relation of the bottom panel of figure 5 in
McLure et al. (2004). We note that using the same $\delta r_{e}$ factor for
all galaxies simply produces a uniform translation of the objects with
respect to the $\log r_{e}$ and $\log I_{e}$ axes and thus affects the
offset but not the slope of the corresponding ZP5 plane. However, for each
value of $\delta r_{e}$ we re-fit a plane (rather than simply adjust the
$\gamma$ plane parameter) in order to account for the uncertainties in
the $\alpha$ and $\beta$ plane parameters.
The best fit is obtained with $\delta r_{e} = 0.39$
(although the reduced-$\chi^{2}$ is fairly constant between 0.3 and 0.6)
and is shown in Figure \ref{evolvedRhalfsFP}.
This is in contrast to a value of $\sim$0.91 expected from the model of
van der Wel et al. (where we consider only the effects of mergers and
exclude the effects of late-type galaxies turning into early-types) but
the upper end of our range
is in good agreement with the model of Khochfar \& Silk (2006) who predict
a value of $\delta r_{e} \sim 0.63$--$0.64$ for the most massive galaxies.
The size evolution implied by $\delta r_{e} = 0.39$ would result in
a mean scalelength for our objects of 37.7 kpc. This is similar to the
size of Brightest Cluster Galaxies (BCGs) in the local universe
(Graham et al. 1996; McLure et al. 2004), which suggests that the host
galaxies of powerful radio galaxies may evolve into BCGs.

However, as discussed above, whilst applying
a uniform size evolution factor to our ZP5 objects can reduce the offset
to the local galaxies, it does not affect the rotation.
This is seen in Figure \ref{evolvedRhalfsFP} where the tilt of
the plane is still not consistent with the local data of
Bettoni et al. (2001). Indeed, whilst the reduced-$\chi^{2}$ obtained
by comparing the size-evolved ZP5 data to the plane of Figure
\ref{evolvedRhalfsFP} is 2.9 (Table \ref{table:normalredcs}),
or 1.2 when we use errors $\geq 15 \%$, when we compare the data of
Bettoni et al. to this plane the reduced-$\chi^{2}$ is 4.8
(or 1.9 with $\geq 15 \%$ errors).
Similarly, fitting a plane to the combined data of Bettoni et al.
and the size-evolved ZP5 data yields a plane 
($\alpha = 1.32$, $\beta = 0.61$, $\gamma = 0.7$) whose
reduced-$\chi^{2}$ values are 4.5, 5.3 and 3.4
when compared to the combined data, the ZP5 data and the data
of Bettoni et al. respectively. Using errors $\geq 15 \%$ these
values reduce to 1.8, 2.3 and 1.3 respectively. We therefore
find that size evolution of the host galaxies by a uniform
factor is not sufficient, in and of itself, to explain the difference
between the $z=0$ and the $z \sim 0.5$ Fundamental Planes, since
it doesn't address the difference in rotation between these two planes.

In order for the tilt to be explained by size
evolution we would require a mass (and size) dependent evolution of the host
galaxies. Once more comparing the ZP5 plane fitting in row one of Table
\ref{table:rhalfsplanes} with the local fitting in row three of Table
\ref{table:bettoniplanes}, and taking $\delta r_{e}$ (as defined above)
to be a function of mass, we find $\delta r_{e} \propto r_{e}^{1.52} M^{-2.11}$
(neglecting any other forms of evolution). This seems more plausible than the
mass-dependent mass-to-light ratio evolution, although the actual evolutionary
factors required for our objects (0.09--86.7, calculated using the same
method) are not. A mass-dependent
size evolution could be driven by environmental effects (where the most
massive galaxies are found only in virialized clusters where mergers are
suppressed), which we will investigate (Herbert et al., in preparation)
using deep multi-band imaging data.

We therefore find that, whilst a mass-dependent size evolution may be the
dominant effect behind the evolution of the Fundamental Plane, it is not
sufficient in and of itself to explain the observed rotation between the
$z \sim 0.5$ and $z = 0$ planes. We suggest that size evolution may combine
with passive evolution and a mass-dependent evolution of the mass-to-light
ratio to produce the observed rotation.

\subsection{A Link with Recent Star Formation?}\label{discussion:sfh}

We next consider the possibility that our powerful ZP5 radio galaxies
already inhabit the local Fundamental Plane, but that the effects of
recent star formation make them appear to lie off it. In other words, is
it possible that our objects that lie below the local Fundamental Plane
could have a bright extended disc of recent star formation that would
increase the observed effective radius?  Conversely, is it possible
that our objects that lie above the local plane could have a bright
nucleocentric region of recent star formation that would decrease the
observed effective radius? If the star formation terminates by $z = 0$
then the observed effective radius would be altered in both cases, possibly
moving the objects back on to the local relation.

\begin{figure}
\centering
\includegraphics[width=0.45\textwidth]{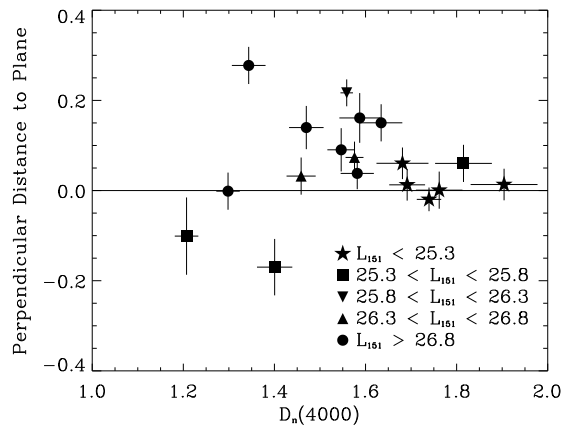}
\caption{\label{Residuals}The residual perpendicular to the
plane of Figure \ref{absoluteBettoniFP} versus D$_{n}$(4000)
for the ZP5 objects ($z \sim 0.5$). The symbols are as used in Figure
\ref{absoluteBettoniFP}.}
\end{figure}

We test this hypothesis using the strength of the 4000\ang\ break for our
objects; the D$_{n}$(4000) indices (taken from Herbert et al. 2010)
can be used as a measure of recent star formation, where younger populations
have smaller 4000\ang\ breaks, as discussed in Herbert et al. (2010).
In Figure \ref{Residuals} we show the residual perpendicular to
the plane of Figure \ref{absoluteBettoniFP} versus D$_{n}$(4000)
for the ZP5 objects. We see that those objects that fall
significantly below the local Fundamental Plane, as well as the two
furthest above it, all have lower D$_{n}$(4000) indices and thus younger
stellar populations. This lends a degree of support to the
recent star formation hypothesis, although we note that some of
the objects that fall furthest from the plane have no evidence of
recent star formation whilst some falling on or near the plane do
have evidence for younger stellar populations.

In order to test the hypothesis further we extract the spectra
for our objects using a range of apertures across the galaxy. We calculate D$_{n}$(4000)
indices in each case and look for any trends between D$_{n}$(4000) 
and extraction radius for each object. However, we find no such trends
within the uncertainties on D$_{n}$(4000) and the amount
of possible contamination discussed in Herbert et al. (2010).
We cannot therefore, with our current data, prove star formation
in extended discs and nucleocentric regions as an explanation for
why some of our objects lie away from the local Fundamental Plane.

\subsection{Correlations with Radio Luminosity}\label{discussion:L151}

Evidence of a correlation between black hole mass and radio
luminosity was first presented by Franceschini, Vercellone
\& Fabian (1998), who showed that both the nuclear and the
total radio fluxes of their sample of nearby radio galaxies
displayed a remarkably tight dependence on black hole mass.
This result has been confirmed by more recent work (e.g.
McLure et al. 1999; Lacy et al. 2001; McLure \& Jarvis 2004)
whilst evidence has also emerged of a connection between
black hole mass and radio loudness (e.g. Laor 2000;
McLure \& Dunlop 2001; McLure \& Jarvis 2004), although other
studies have found no evidence of such connections
(e.g. Ho 2002; Woo \& Urry 2002).

In order to investigate the possible link between radio luminosity
and black hole mass for the objects in our sample we
calculate the black hole mass from $\sigma$ using
the $M_{BH}$-$\sigma$ relation for elliptical galaxies published
by G\"ultekin et al. (2009) and propogate errors in the standard
fashion by assuming symmetric errors on $\log \sigma$ as above.
It is important to note that, whilst the
majority of the previous studies have used the high-frequency
(5-GHz) radio luminosity, we use the extended low-frequency
(151-MHz) radio luminosity. This may be significant, since
the 151-MHz radio luminosity is closely connected with the
time-averaged kinetic energy of the jets (e.g. Rawlings \&
Saunders 1991) and the effects of beaming are smaller on
the 151-MHz radio luminosity than on the 5-GHz radio luminosity
(e.g. Jarvis \& McLure 2002).

\begin{figure}
\centering
\includegraphics[width=0.45\textwidth]{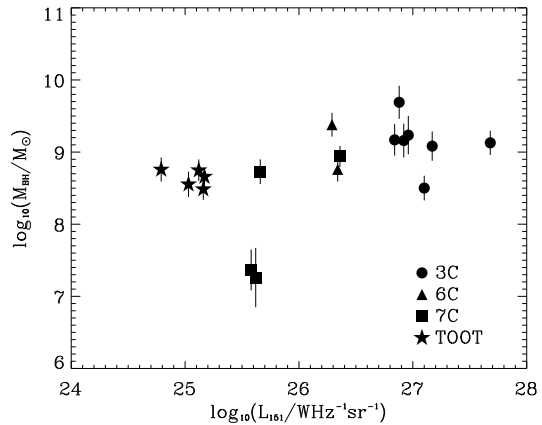}
\caption{\label{L151Mbh}The variation of the logarithm of the
151-MHz radio luminosity, $L_{151}$, with black hole mass for our ZP5 objects ($z \sim 0.5$).
Circles represent 3C galaxies, 6C galaxies are represented
by triangles, 7C galaxies by squares, and TOOT objects
by stars.}
\end{figure}

In Figure \ref{L151Mbh} we show $L_{151}$ versus black hole mass.
Figure \ref{L151Mbh} provides evidence in support of a link
between radio luminosity and black hole mass:
using the Spearman rank correlation coefficient we find a positive
correlation between the $L_{151}$ radio luminosity and $M_{BH}$
at a significance level of 97\%. This is
in good agreement with the $L_{151}$-$M_{BH}$ relation for the whole
ZP5 sample found at a significance level of 97\% by McLure et al.
(2004) where the black hole masses were estimated via the
$M_{BH}$-$M_{Bulge}$ relation of McLure \& Dunlop (2002). However,
in contrast to McLure et al. (2004), who find that the significance of
the correlation is increased to 99.5\% by excluding the TOOT
objects, we find that excluding the TOOT objects decreases the
correlation significance to 84\%. However, this could be purely
due to the decrease in the number of objects we use for our analysis;
24 compared to the full sample of 41 used in the work of McLure et al.
(2004). Additional spectroscopy on the remaining 27 radio galaxies
would be needed to confirm this.

\begin{figure}
\centering
\includegraphics[width=0.45\textwidth]{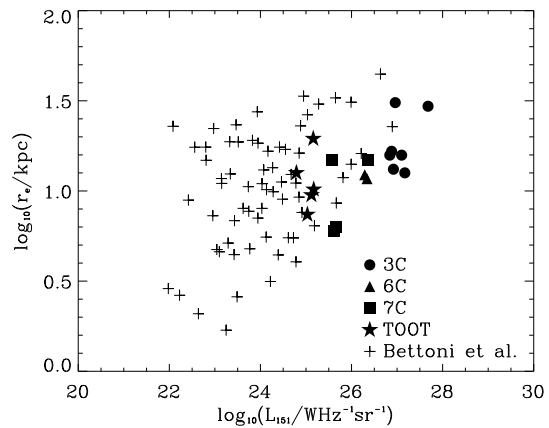}
\caption{\label{L151re}The variation of the logarithm of the
151-MHz radio luminosity, $L_{151}$, with $r_{e}$ for our ZP5 objects ($z \sim 0.5$). The symbols
are as used in Figure \ref{L151Mbh}, with the addition of the data of Bettoni
et al. (2001; $z \sim 0.03$), shown by crosses.}
\end{figure}

We also find a positive correlation between $L_{151}$ and $r_{e}$
(Figure \ref{L151re}). Applying the Spearman rank correlation coefficient
yields a significance of 98\%. It is interesting to note that the
TOOT objects (triangles) --- FRI type objects with 
$L_{151} < 10^{25.3}$ W~Hz$^{-1}$~sr$^{-1}$ --- do not appear to follow
this relation; for the TOOT objects $r_{e}$ appears to be largely
independent of $L_{151}$. This is consistent with our findings for
the Fundamental Plane, where the low-luminosity radio sources are
consistent with being able to passively evolve onto the local relation.


\section{Conclusions}\label{conclusions}

In this paper we have presented deep spectroscopic observations of
24 powerful radio galaxies at $z \sim 0.5$. We have used a direct
fitting procedure to determine the velocity dispersions
of our objects and combined these with the host galaxy data from
McLure et al. (2004) to study the Fundamental Plane.

\begin{itemize}

\item We have shown that the FRI type objects in our sample, once
corrected for passive evolution, lie on the Fundamental Plane
inhabited by the local radio galaxies of Bettoni et al. (2001).
The FRII type objects in our sample do not lie on the local
plane. However, whilst the majority of the sample of Bettoni et al. are
lower radio luminosity (FRI type) objects, the FRII type objects
in their sample are consistent with the local plane. Furthermore,
our FRI type objects and our FRII type objects appear able to share
the same plane at $z \sim 0.5$. We therefore suggest that
radio galaxies with lower and higher
radio luminosities may share the same Fundamental Plane, but that there is
substantial evolution in this plane between $z=0$ and $z \sim 0.5$.

\item One explanation for the tilt observed between the $z = 0$ and
the $z \sim 0.5$ Fundamental Planes is evolution of the host galaxies.
We consider passive evolution, a mass-dependent evolution
in the mass-to-light ratio and an evolution in the size of the host galaxy,
but find that none of these effects can, by themselves, plausibly explain the
observed rotation. We suggest, however, that some combination of all three effects,
with size evolution as the dominant factor, may be sufficient
to explain the difference between the planes. Indeed,
indirect evidence for size evolution for the high-luminosity sources comes from
the significant correlation (98\%) between the effective radius and the radio
luminosity for these sources, whereas the FRI sources do not appear to
align with such a relation. 

\item We also consider the possibility that our radio galaxies do, in fact,
lie on the local Fundamental Plane (once passive evolution has been
corrected for), but that a bright extended disc or
nucleocentric region of recent star formation alters the observed
effective radius, thus causing some of our objects to appear to lie
off the local relation. Using the D$_{n}$(4000) index as an indicator
of recent star formation, we see some hints that this may be having an 
effect, but the evidence is far from conclusive.

\item We find evidence at the 97\% level of a correlation between galaxy
velocity dispersion and the radio luminosity, suggesting that radio
luminosity scales with black-hole mass. This is in line with previous
work on this sample using the host galaxy luminosity as a proxy for black-hole
mass (McLure et al. 2004). Unlike McLure et al., however, we do not find
that this correlation becomes stronger in the absence of the low-luminosity
(FRI-type) radio sources. We attribute this to the lack of objects in the
present study (24) compared with the analysis of the full sample of 41
in McLure et al..

\end{itemize}

In future work (Herbert et al. in preparation) we will use deep
multi-band imaging data to investigate the environments of the full
$z \sim 0.5$ sample and how they relate to the power of the
AGNs and the masses and star formation histories of the host galaxies.

\section*{ACKNOWLEDGEMENTS} 
{\footnotesize}
We thank the referee, Brant Robertson, for his useful comments.
PDH thanks the UK STFC for a studentship. MJJ ackowledges the
support of an RCUK fellowship. RJM and JSD acknowledge the support
of the Royal Society through a University Research Fellowship and
a Wolfson Research Merit award respectively.
The William Herschel Telescope is operated on the
island of La Palma by the Isaac Newton Group in the Spanish
Observatorio del Roque de los Muchachos of the Instituto de Astrofisica
de Canarias.
Based on observations obtained at the Gemini Observatory
(programs GN-2008B-Q-103 \& GN-2009A-Q-105),
which is operated by the Association of Universities for
Research in Astronomy, Inc., under a cooperative agreement
with the NSF on behalf of the Gemini partnership: the
National Science Foundation (United States), the Science
and Technology Facilities Council (United Kingdom), the
National Research Council (Canada), CONICYT (Chile), the
Australian Research Council (Australia), Minist\'erio da
Ci\^encia e Tecnologia (Brazil) and Ministerio de Ciencia,
Tecnolog\'ia e Innovaci\'on Productiva (Argentina).

{}

\section*{APPENDIX}

\begin{figure*}
\centering
\includegraphics[width=0.95\textwidth]{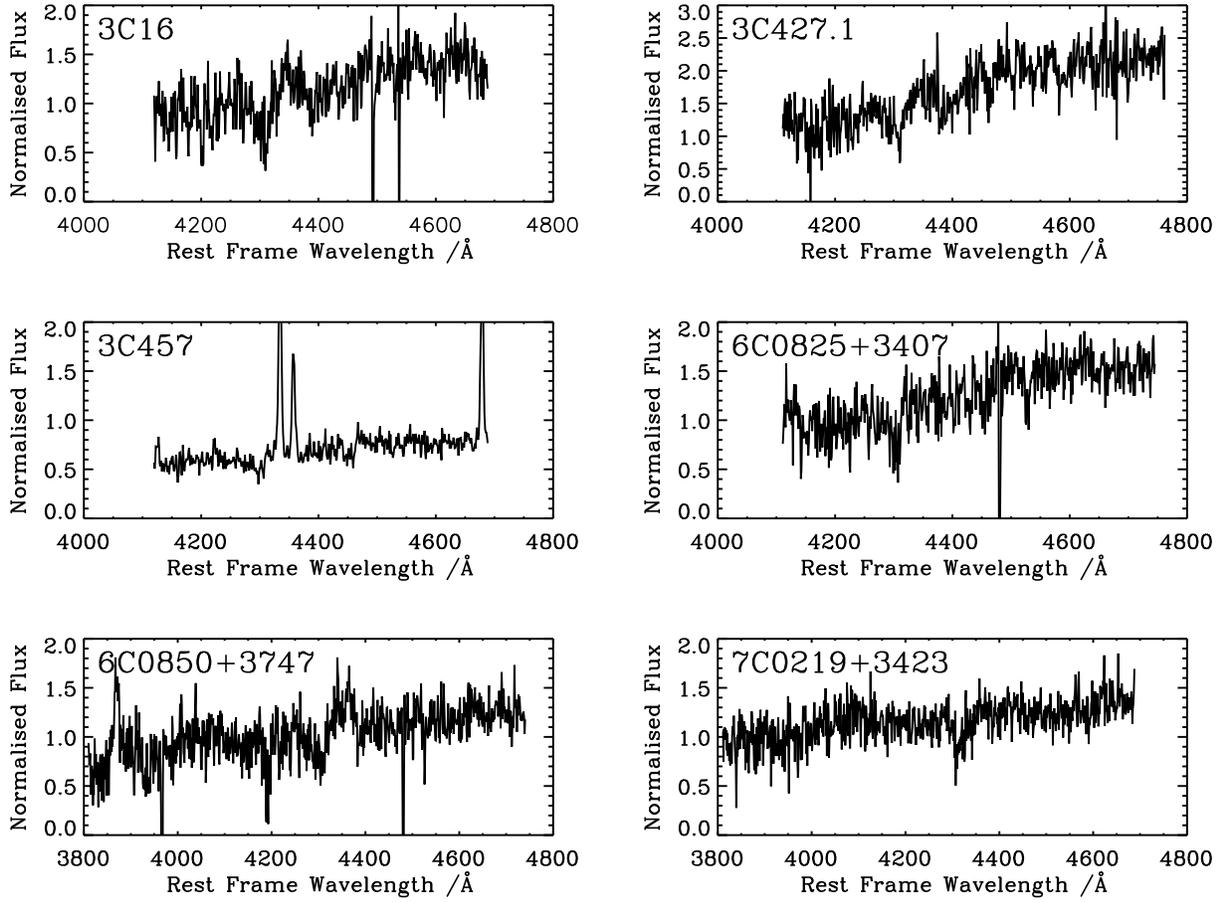}
\caption{\label{no-fit-spectra}Spectra of the six
objects for which we were unable to obtain a reliable
measurement of the velocity dispersion. 3C457 was rejected
due to a combination of the signal-to-noise
ratio and emission lines from the AGN. The remainder were
rejected on the basis of their signal-to-noise ratios alone
(see Section \ref{fittingprocedure}).}
\end{figure*}

In Figure \ref{no-fit-spectra} we show the spectra of the
six objects for which we were unable to obtain a reliable
measurement of the velocity dispersion (Section \ref{fittingprocedure}).

\end{document}